\documentclass[reprint,
superscriptaddress,
 amsmath,amssymb,
 aps,prl,longbibliography
]{revtex4-1}

\usepackage{graphicx}
\usepackage{dcolumn}
\usepackage{bm}
\usepackage{color}
\usepackage{xcolor}

\usepackage{amsmath}

\definecolor{airforceblue}{rgb}{0.36, 0.54, 0.66}

\begin{document}

\preprint{APS/123-QED}

\title{Cross-band versus intra-band pairing in superconductors:\\ signatures and consequences of the interplay}

\author{A. A. Vargas-Paredes}%
\affiliation{School of Pharmacy, Physics Unit, Universit\`a di Camerino, 62032 Camerino, Italy}
\affiliation{Department of Physics, University of Antwerp, Groenenborgerlaan 171, B-2020 Antwerp, Belgium}
\author{A. A. Shanenko}
\affiliation{Universidade Federal de Pernambuco, Av. Prof. Luiz Freire, s/n, 50670-901 Recife-PE, Brazil}
\author{A. Vagov}
\affiliation{Institut f\"ur Theoretische Physik III, Bayreuth Universit\"at, Bayreuth, D-95440, Germany}
\author{M. V. Milo\v{s}evi\'c}
\affiliation{Department of Physics, University of Antwerp, Groenenborgerlaan 171, B-2020 Antwerp, Belgium}
\affiliation{School of Pharmacy, Physics Unit, Universit\`a di Camerino, 62032 Camerino, Italy}
\author{A. Perali}
\affiliation{School of Pharmacy, Physics Unit, Universit\`a di Camerino, 62032 Camerino, Italy}
\date{\today}

\begin{abstract}
We analyze the paradigmatic competition between intra-band and cross-band Cooper-pair formation in two-band superconductors, neglected in most works to date. We derive the phase-sensitive gap equations and describe the crossover between the intraband-dominated and the crossband-dominated regimes, delimited by a ``gapless'' state. Experimental signatures of crosspairing comprise notable gap-splitting in the excitation spectrum, non-BCS behavior of gaps versus temperature, as well as changes in the pairing symmetry as a function of temperature. The consequences of these findings are illustrated on the examples of MgB$_2$ and Ba$_{0.6}$K$_{0.4}$Fe$_2$As$_2$.
\end{abstract}

\maketitle

Multiband superconductivity is known to promote novel quantum phenomena of great fundamental importance and versatility \cite{milo_perali}. Among recent examples are optically excited collective modes in multiband MgB$_2$ \cite{giorgianni}, the emergent phenomena at the BCS-BEC crossover in FeSe \cite{hanaguri}, and at oxide intefaces \cite{bergeal}. Strong scientific appeal of multiband superconductivity stems from its pronounced tunability. External pressure, lattice strain effects, gating, chemical doping, photo-induction, quantum confinement and surface effects are all able to move and change the band dispersions and the position of the chemical potential with respect to Lifshitz transitions \cite{bergeal,li, wx_li, costanzo, continenza, porta}, where superconducting properties can radically change. 

\begin{figure}%
\includegraphics[width=\linewidth]{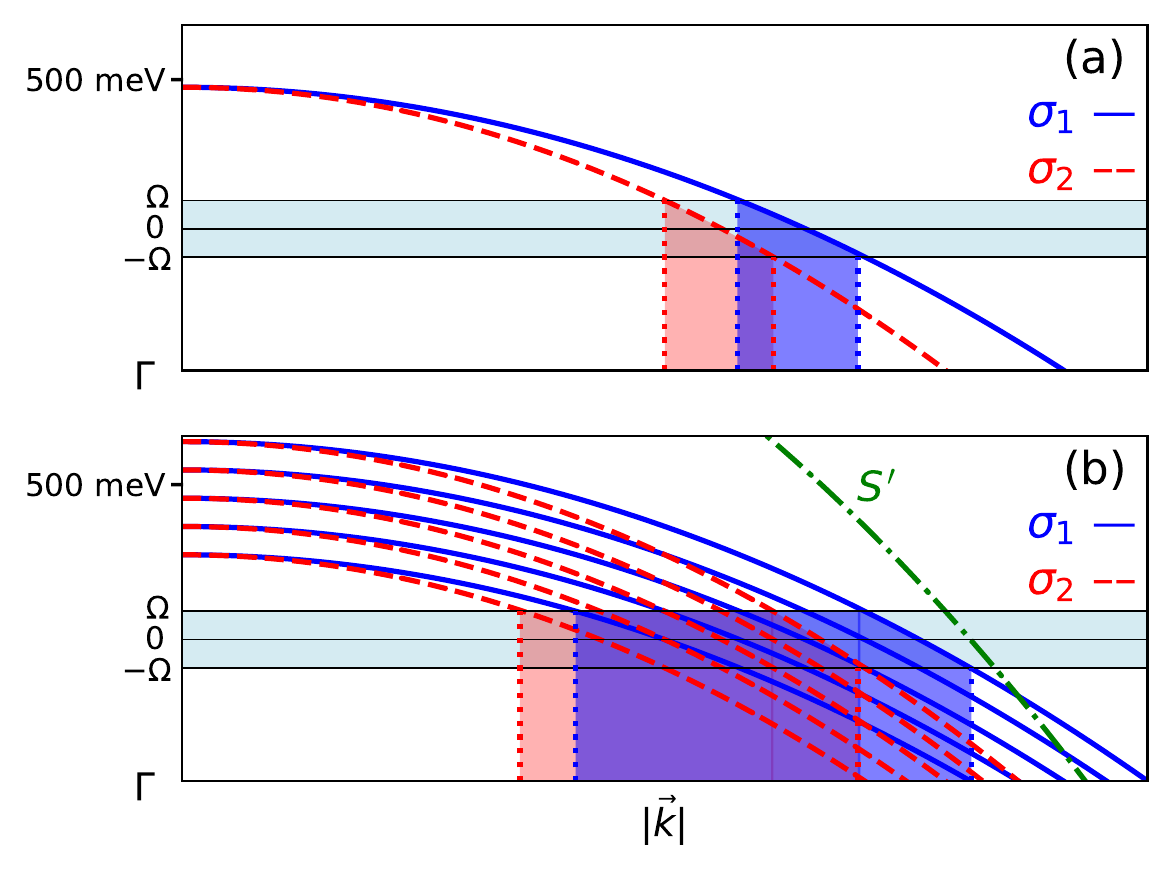}%
\caption{The relevance of crosspairing is illustrated based on the band structure of (a) bulk MgB$_2$ \cite{ponce} and (b) 6-monolayer MgB$_2$ \cite{jonas}. Only $\sigma$ bands close to the $\Gamma$ point of the Brillouine zone are shown, with chemical potential $\mu=500$ meV and energy scale of the pairing $\Omega=75$ meV. In (b), each interior monolayer contributes a pair of hole-like bands $\sigma_1$ and $\sigma_2$, and the surface band is denoted by $S'$. The (purple) overlapping shadows project the momentum states where cross-band pairing between opposite momenta states among the $\sigma$ bands is feasible.}\label{fig1}
\end{figure}

To date, multiband electronic structure is proven to be of crucial importance in rather versatile superconducting systems, such as MgB$_2$ \cite{akimitsu}, iron-based compounds \cite{selective_mott, nemat_pairing_orbital_select, orbital_selec1, orbital_selec2, odd_freq1, ding1}, superconducting nanostructures \cite{bianconi97, valletta97, cross_arkady, sc_nanoribbon, surface_step}, 2D electron gases at interfaces \cite{valentinis, mohanta, thais1}, metal-organic superconductors \cite{chain_molecule, met_org, possible_fano}, etc. In such multiband superconductors, the pairing interaction and the proximity/hybridization of two or more bands can result in the formation of Cooper pairs with electrons originating from different bands, a phenomenon termed  ``cross-band pairing" or simply ``crosspairing". This pairing is to be distinguished from the Josephson-like pair transfer between the intraband condensates, which is usually taken as their sole coupling in multiband superconductors. Crosspairing and intraband pairing are intuitively competitive, therefore it is necessary to understand their interplay qualitatively and quantitatively, together with associated changes in physical properties and observables. Such understanding is far from established, as crosspairing and its competition with intraband pairing were predominantly neglected in the studies to date. In superfluid systems with at least two fermionic species, the partially overlapping bands at the Fermi level are prone to crosspairing, as discussed in Refs. \onlinecite{wilczek1, wilczek2}. In superconductors, the hybridization of multiple bands close to the Fermi level is favorable for cross-band pair formation. This occurs in the iron-based superconductors (FeSCs) which present hybridized orbitals \cite{amoreo1, amoreo2}, cuprates with the hybridization of $d_{x^2-y^2}$  and $d_{z^2}$ orbitals \cite{matt2018, jamil_tahir}, and also in the heavy-fermion compounds, where crosspairing between electrons with $f$ and $d$ orbital character has been considered \cite{dolgov}. However, even without hybridization, the plain proximity of multiple bands can facilitate crosspairing, as illustrated in Fig. \ref{fig1} for bulk and atomically-thin MgB$_2$. 

In this Letter, we examine the interplay between intra- and cross-band pairing in two-band superconductors and its experimental signatures. We reformulate the mean-field equations for the superconducting order parameter, going beyond the Suhl, Matthias and Walker (SMW) extension of the BCS theory \cite{suhl_matthias}. This results in an extended self-consistent and phase-dependent set of equations for the several components of the order parameter, with strongly hybridized excitation spectra. The mean-field Hamiltonian including both intraband and cross-band pairing reads:

\begin{eqnarray}
H &=& \sum_{i,\textbf{k}, \sigma}\epsilon_{i}(\textbf{k})c_{i,\textbf{k}\sigma}^\dag c_{i,\textbf{k}\sigma} + H_{int}, \label{ham} \\
H_{int} &=& \sum_{i,j}\sum_{\textbf{k}}\left[ \Delta_{ij}(\textbf{k})c^\dagger_{i,\textbf{k}\uparrow}c^\dagger_{j,-\textbf{k}\downarrow} + h.c. \right],
\end{eqnarray}
where $i,j=1,2$ represent the band index, and $\sigma=\uparrow,\downarrow$ the spin. Here $\Delta_{ij}({\bf k})=\sum_{k,l=1,2}\sum_{\textbf{k}'}V_{ij,kl}(\textbf{k},\textbf{k}')\left<c_{k, -\textbf{k}\uparrow }c_{l,\textbf{k}'\downarrow}\right>$ are the pairing amplitudes, and $\epsilon_{i}(\textbf{k})$ is the band-dependent kinetic energy of the electrons. We note that Eq. \eqref{ham} resembles the Hamiltonian of a two-band system with hybridization upon the change from orbital to the band basis \cite{amoreo2}. The full \textbf{k}-dependent form of the interaction matrix is given by $V_{ij,kl}(\textbf{k},\textbf{k}') = -g_{ij,kl}\Theta(\Omega - |\zeta_i(\textbf{k})|)\Theta(\Omega - |\zeta_j(\textbf{k})|)\Theta(\Omega - |\zeta_k(\textbf{k}')|)\Theta(\Omega - |\zeta_l(\textbf{k}')|)$, where $\Omega$ is the average energy scale of the effective interaction, and $\zeta_i(\textbf{k}) = \epsilon_i(\textbf{k}) -\mu$ with chemical potential $\mu$. In $g_{ij,kl}=
\begin{pmatrix}
g_{11,11} & g_{11,22} & g_{11,(12)} \\ 
g_{22,11}&g_{22,22} & g_{22,(12)} \\ 
g_{(12),11} & g_{(12),22} & g_{(12),(12)}
\end{pmatrix}$,
the upper left $2\times2$ inner matrix corresponds to the well established SMW case \cite{suhl_matthias}, and the third row and column include the crosspairing (where (12) indicates symmetrization under given indices, so that e.g., $g_{(12),(12)}=g_{12,12}+g_{21,21}$). In the interaction matrix the effective attraction between electrons is given by its diagonal elements, and the off-diagonal ones describe the Josephson-like coupling between intraband and cross-band condensates. 

In what follows, we simplify our indices as $11\equiv 1$, $22\equiv 2$ and $(12)\equiv 3$. Next, we use the Gor'kov Green's functions formalism to obtain the pair amplitude equations \cite{korko_palist, cross_arkady}. In momentum space the two excitation spectra without crosspairing ($i=1,2$) are $\varepsilon_i^2=\zeta_i^2+\left|\Delta_{i}\right|^2$ and the pair amplitudes are given by $\Delta_{i}(\textbf{k})=\left|\Delta_{i}\right|e^{i\varphi_{i}}\Theta (\Omega-\left|\zeta_i(\textbf{k})\right|)$, where $\varphi_i$ is the phase of the pair amplitude. 

The crosspairing pair amplitude $\Delta_{3}$ hybridizes the energy spectra of the two BCS-like excitation branches: 
\begin{eqnarray}
E_{\pm}(\theta)&=&\sqrt{\frac{1}{2}\left(\varepsilon_1^2+\varepsilon_2^2+2\left|\Delta_{3}\right|^2\pm b(\theta)\right)},\label{bands2}\\
b(\theta)&=&\sqrt{\left(\varepsilon_1^2-\varepsilon_2^2\right)^2+4\left|\Delta_{3}\right|^2r(\theta)}, \label{bconst}
\end{eqnarray}
where $r=\left(\zeta_1-\zeta_2\right)^2+\left|\Delta_{1}\right|^2+\left|\Delta_{2}\right|^2+2\left|\Delta_{1}\right|\left|\Delta_{2}\right|\cos\theta$ and $\theta=2\varphi_{3}-\varphi_{1}-\varphi_{2}$. We emphasize here that the angle $\theta$ will introduce new degrees of freedom in our system depending on the combination of the couplings, as will be shown later. The excitation gaps $\Delta_{\pm}(\theta)$ coincide with the minimum energy of the excitation branches $E_{\pm}(\theta)$. These are the two gaps $\Delta_{\pm}$ present in the density of states (DOS), however these gaps no longer correspond to the energy needed to break intraband Cooper pairs (as is conventionally the case). Instead, they describe the energy needed to disallow \textit{either} intra- or cross-band pairing. 

The self-consistent equations for the pair amplitudes are given by:
\begin{equation}
\Delta_{i}=\frac{1}{2}\sum_{j}g_{ij}\int{\frac{d^3k}{(2\pi)^3}}\Delta_{j}\left[\chi_{j}^+f\left(E_+\right)+\chi_{j}^-f\left(E_-\right)\right],\label{mgap}
\end{equation}
where $f(E)=\frac{1}{2E}\tanh\left(\frac{\beta E}{2}\right)$, $\chi_{i}^{\pm}=1\pm\frac{1}{b(\theta)}\chi_{i}$, $\chi_{1(2)}=\varepsilon_{1(2)}^2-\varepsilon_{2(1)}^2+2\left|\Delta_{3}\right|^2\left(1+\left|\Delta_{2(1)}\right|e^{i\theta}/\left|\Delta_{1(2)}\right|\right)$, $\chi_{3}=\left(\zeta_1-\zeta_2\right)^2+\left|\Delta_{1}\right|^2+\left|\Delta_{2}\right|^2+2\left|\Delta_{1}\right|\left|\Delta_{2}\right|e^{-i\theta}$, and $\beta=1\big/k_BT$. Note that these pairing amplitudes (i.e. the order parameters in the problem) {\it do not correspond} to the measurable gaps $\Delta_\pm$.

\begin{figure}%
\includegraphics[width=\linewidth]{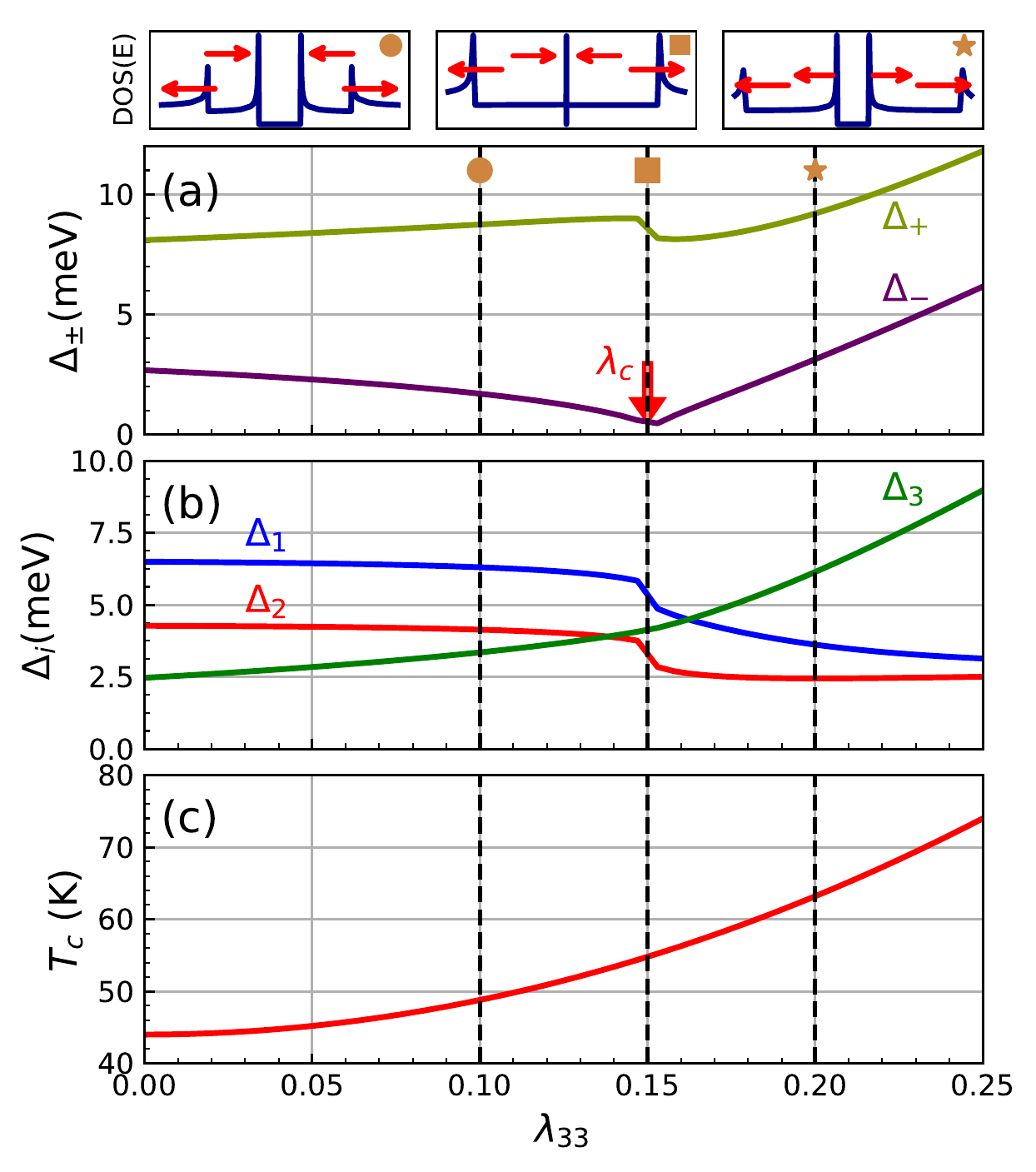}%
\caption{Effect of crosspairing for the same phase. Excitation gaps (a) with their corresponding pair amplitudes (b) as function of $\lambda_{33}$ at $T=4.2$ K. The three miniplots above (a) show the density of states for $\lambda_{33}=$ $0.1$, $0.15$, and $0.2$, illustrating the behavior in the intraband-dominated regime, gapless state, and the crosspairing dominated regime, respectively. (c) Mean-field critical temperature versus $\lambda_{33}$.}%
\label{fig2}%
\end{figure}

Before solving the above formalism to reveal new physics brought by crosspairing, we introduce parabolic bands and dimensionless effective couplings, $\lambda_{ij}=g_{ij}N_j(0)$, where $N_{j=1,2}(0)$ is the band-dependent density of states and $N_3(0)=N_1(0)+N_2(0)$. We start by solving Eq. (\ref{mgap}) when all couplings $\lambda_{ij}$ are positive and with the same phase, i.e. $\theta=0$. To visualize the effect of crosspairing we fix all parameters but $\lambda_{33}$: $E_F=200$ meV, $\Omega=30$ meV, $\lambda_{11}=0.4$, $\lambda_{22}=0.3$, $\lambda_{ij, i\neq j}=0.05$. In Figs. \ref{fig2}(a) and \ref{fig2}(b), we show the excitation gaps and all three pairing amplitudes at $4.2$ K. As crosspairing coupling $\lambda_{33}$ is increased, the two excitation gaps $\Delta_{+}$ and $\Delta_{-}$ are split further apart: increasing $\Delta_{3}$ strengthens $\Delta_{+}$ and suppresses $\Delta_{-}$, up to a characteristic value $\lambda_{33}=\lambda_{c}$ (roughly half the average of $\lambda_{11}$ and $\lambda_{22}$). This characteristic value marks the maximal competition between the intraband and the crossband pairing channels and separates the two regimes: the intraband-dominated regime (IDR) for $\lambda_{33}<\lambda_{c}$, and a crosspairing-dominated regime (CDR) for $\lambda_{33}>\lambda_{c}$. In the CDR, both gaps increase at the same rate, similarly to the one-band scenario. Therefore the CDR describes a two-gap system which is characterized by a sole order parameter $\Delta_{3}$, while the intraband pair amplitudes participate only passively, by proximity \cite{roditchev_prox_effect, roditchev_prox_effect2}. Fig. \ref{fig2}(c) shows that superconducting critical temperature $T_c$ increases with $\lambda_{33}$ faster than expected considering the range of values of $\lambda_{33}$ alone.  

\begin{figure}%
\includegraphics[width=0.8\linewidth]{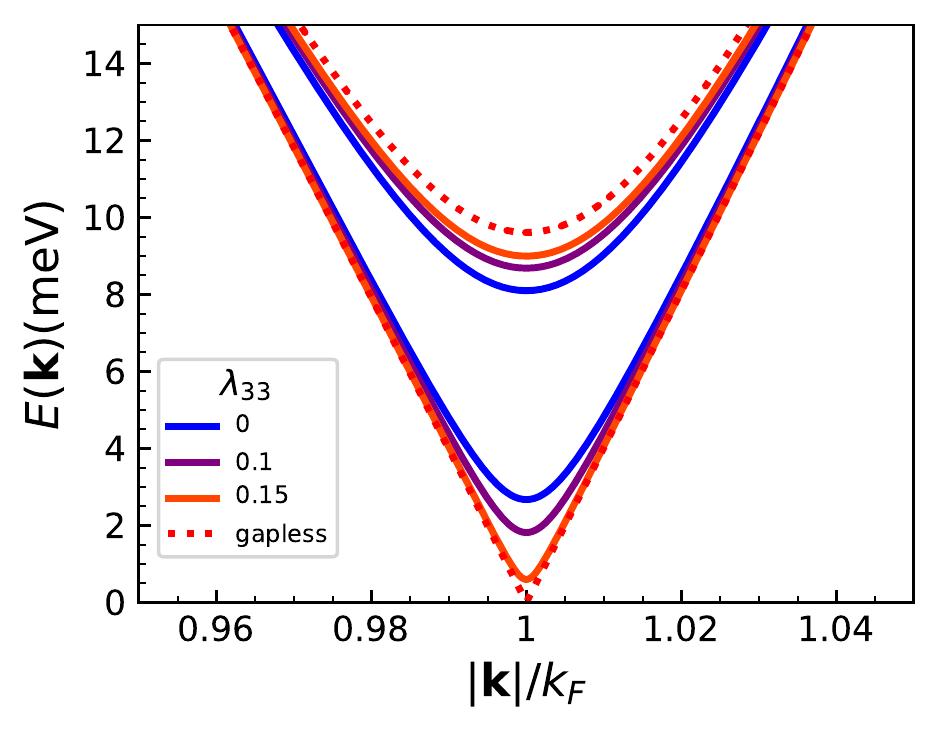}%
\caption{Energy dispersions corresponding to Fig. \ref{fig2}, for $\lambda_{33}$ increased towards the gapless state.}
\label{fig3}%
\end{figure}

In the miniplots above Fig. \ref{fig2}(a), we show the density of states (as a measurable quantity in STM/STS) for the IDR, CDR as well as for the crossover point $\lambda_{33}=\lambda_c$. Note that in the latter situation the inner coherence peak approaches zero energy, and may disappear at exactly zero for a favorable combination of parameters. That case would mark a {\it gapless regime}, where the weaker gap is no longer directly detectable, but must play a role in all observables in e.g., applied magnetic field or transport measurements. In such a state, superconducting gaps extracted from the tunneling spectra of STM would no longer coincide with the ones extracted from low-temperature ARPES \cite{dama} using normal-state band structure as a reference. Moreover, the lowest energy excitation branch exhibits linear V-shaped dispersion in the gapless state (see Fig. \ref{fig3}). Such a multiband system has a peculiar multicomponent composition, with the coexistence of a large-gap condensate and the in-gap states having a free-particle character. This leads to a finite DOS at low energies, and radically changed temperature dependence of all superconducting properties with respect to the gapped state. One concludes that such a gapless state, induced by crosspairing, is a unique feature of multiband superconductors worthy of further investigation. 

\begin{figure}%
\includegraphics[width=\linewidth]{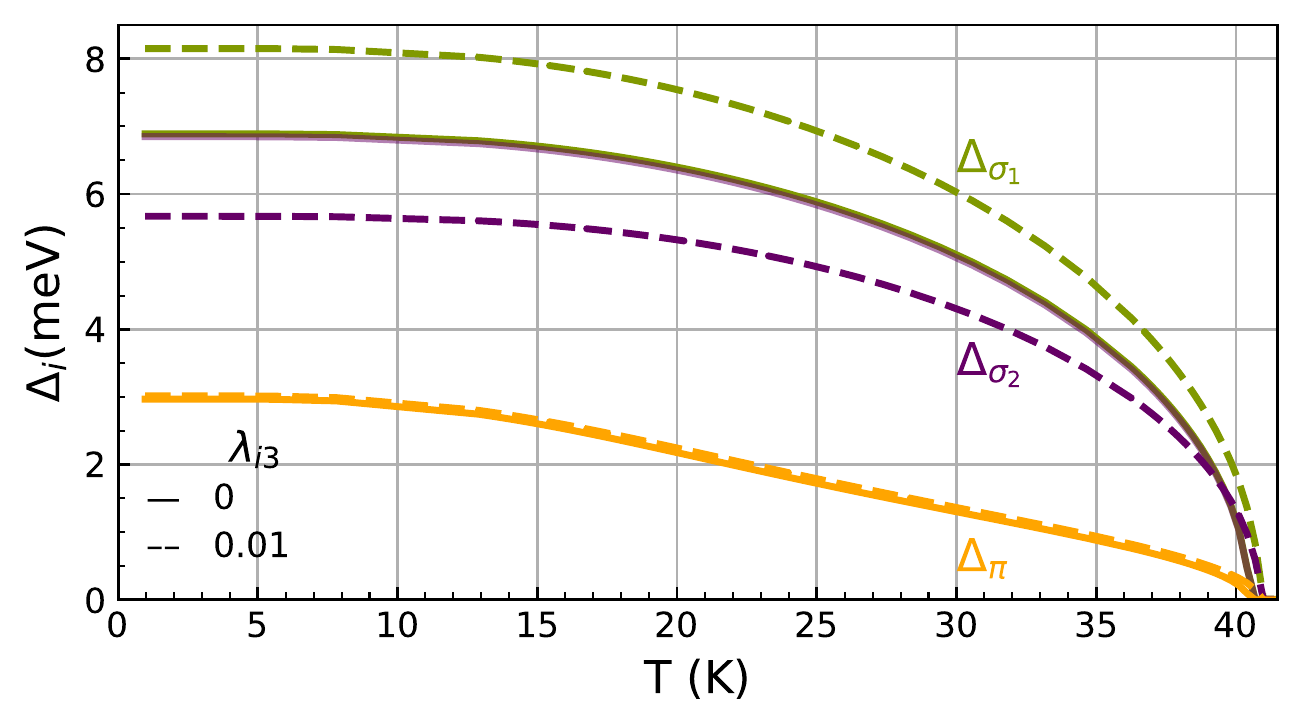}%
\caption{Superconducting gaps of bulk MgB$_2$ as a function of temperature, for intraband pairing only (solid lines), and in the presence of weak crosspairing (dashed lines).}
\label{fig4}%
\end{figure}

To quantify the effects of crosspairing, it is instructive to take the example of the best known two-gap superconductor MgB$_2$ \cite{soumamgb2}. This superconductor has four contributing bands, two $\sigma$-bands for the stronger gap and two $\pi$-bands for the weaker one. The distance of two $\sigma$-bands in the vicinity of the Fermi level is approximately $75$ meV [see Fig. \ref{fig1}(a)]. Taking the parameters $\mu=500$ meV and $\Omega=75$ meV from Refs. \cite{ponce, kuzmichev}, we consider the crosspairing between the $\sigma$-bands, with the coupling matrix
\begin{equation}
\lambda_{ij}=\begin{pmatrix}
0.275 & 0.032 & \lambda_{i3} & 0.032\\ 
0.032 & 0.274 & \lambda_{i3} & 0.032\\ 
\lambda_{i3} & \lambda_{i3} & 0.1 & 0.01\\ 
0.01 & 0.01 & 0.01 & 0.22
\end{pmatrix}. \label{m_mgb2}
\end{equation}
The above matrix is asymmetric because of different DOS associated with each band. $\lambda_{i4}$ is the coupling to the $\pi$ bands, and third column and row correspond to the coupling to the crosspairing channel, with $\lambda_{i3}$ as a free (small) parameter. Other coupling constants are taken from literature, and yield the experimentally verified gaps of MgB$_2$ ($\approx 7$ and 3 meV) in absence of crosspairing ($\lambda_{i3}=0$, see Fig. \ref{fig4}). Even a small $\lambda_{i3}=0.01$ yields a 2 meV split of the two $\sigma$ gaps and a $1$ K increase in $T_c$. This gives confidence that crosspairing effects, even if seemingly small, can lead to significant modifications of the gap spectrum without significantly changing $T_c$. That in turn calls for revisiting of theoretical approaches, e.g., to include crosspairing in anisotropic Eliashberg calculations even for materials that seemed previously well described \cite{choi_mgb2, aperis_mgb2}, as well as revisiting the available experimental data (bearing in mind the non-equivalence between $\Delta_\pm$ and the pairing amplitudes in presence of crosspairing). Conducting more refined ARPES measurements (e.g., in case of crystalline MgB$_2$, on two $\sigma$-bands separately) can provide evidence for the gap splitting caused by crosspairing.

\begin{figure}%
\includegraphics[width=\linewidth]{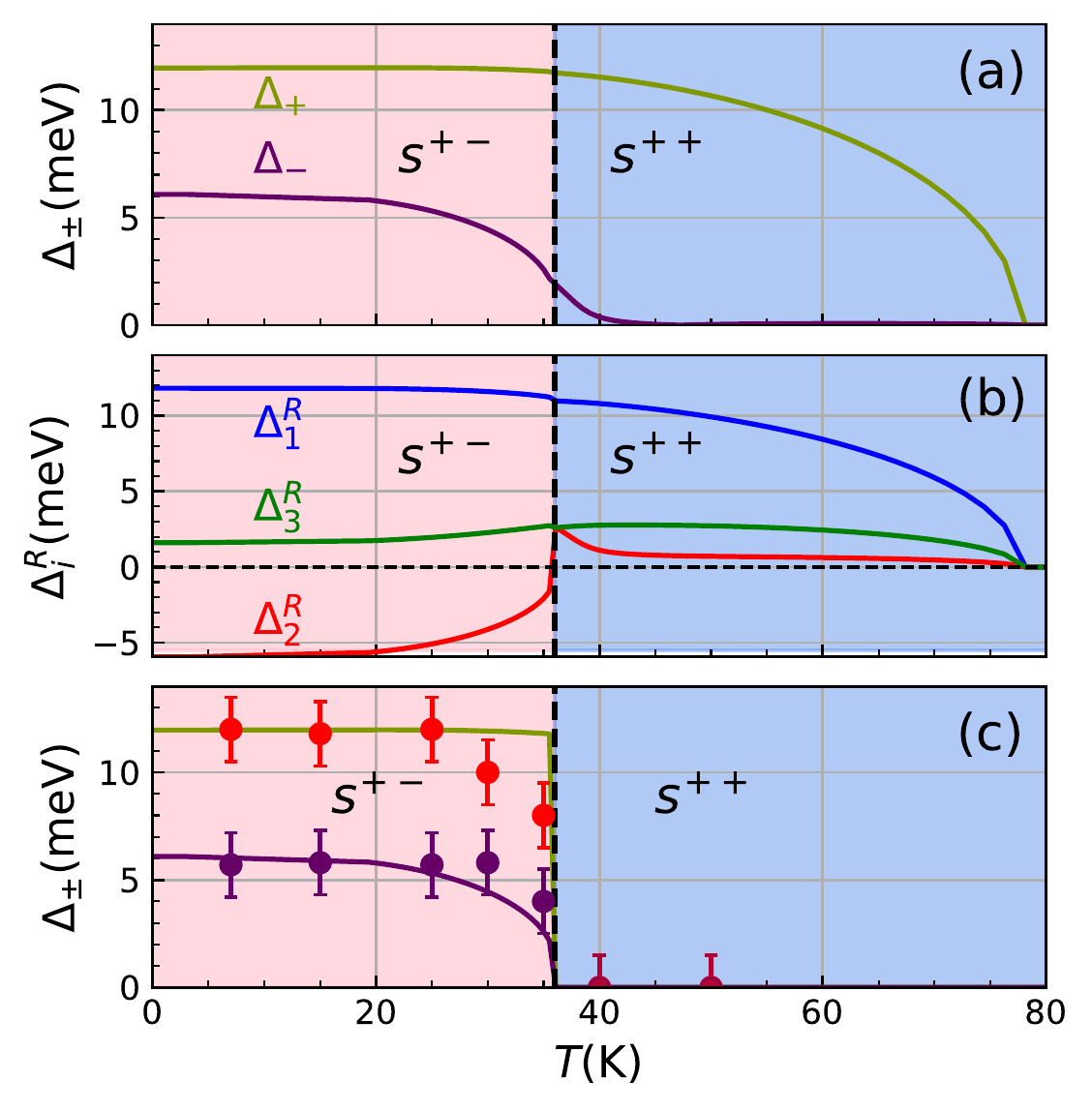}%
\caption{(a) Excitation gaps $\Delta_{\pm}$, and (b) real part of the pair amplitudes $\Delta^R_{i}$ as a function of temperature, for parameters of $(\alpha,\beta)$ bands in (BaK)Fe$_2$As$_2$, with nominal $s^{+-}$ antiphase and in presence of crosspairing. For parameters given in the text, $s^{+-}\rightarrow s^{++}$ transition is found at 36 K, corresponding to the critical temperature of the measured gaps at the $\Gamma$-point of Ba$_{0.6}$K$_{0.4}$Fe$_2$As$_2$ \cite{ding1}. Panel (c) superimposes the theoretical data of (a) on experimental data of Ref. \cite{ding1}, highlighting their agreement in case $s^{+-}$ antiphase is protected.}
\label{fig5}%
\end{figure}

Last but not least, we discuss the phase-frustrated solutions of Eq. (\ref{mgap}), with non-zero angle $\theta$. For example, in the family of FeSCs one can have two cases where a non trivial phase difference is present. The first is the conventional $s^{+-}$ case, which contemplates a $\pi$-phase difference between electron-like and hole-like pair amplitudes \cite{mazin}. The second is the orbital antiphase $s^{+-}$ case, with a $\pi$-phase difference between bands of the same type (electron-like or hole-like), as reported in the optimally doped (BaK)Fe$_2$As$_2$ ($T_c=36$ K) \cite{xiaoli, pzhang, yin2014}. This compound presents two hole-like bands ($\alpha$, $\beta$) stemming from two nested Fermi sheets at $\Gamma$-point, and two electron-like bands ($\gamma$, $\delta$) stemming from two nested Fermi sheets at the M-point. The proximity of both pairs of bands to the Fermi level and the smallness of their interband distance justifies the assumption of crosspairing between bands $\alpha$ and $\beta$ or $\gamma$ and $\delta$. To identify the emergent effects, we will consider the effect of crosspairing only between $\alpha$ and $\beta$ (assuming similar consequences for crosspairing between $\gamma$ and $\delta$). We take the interband distance between $\alpha$ and $\beta$ as $10$ meV and the Fermi level at $\mu=50$ meV, following Ref. \cite{ding2}. To obtain the gaps ($\Delta_\pm$) as measured in low-temperature experiments of Ref. \onlinecite{ding1} ($\approx12.4$ and 6.2 meV extrapolated to $T=0$), we take for the coupling matrix:
\begin{equation}
\lambda_{ij}=\begin{pmatrix}
0.51 & \lambda_{12} & \lambda_{13}\\ 
0.5\lambda_{12} & 0.39 & \lambda_{13}\\ 
0.5\lambda_{13} & 0.5\lambda_{13} & 0.25
\end{pmatrix}.\label{m_fe2as2}
\end{equation}
Here $\lambda_{12}$ is taken negative, which is the standard way to obtain the sign change in the band-dependent order parameters (as reported in Ba$_{0.6}$K$_{0.4}$Fe$_2$As$_2$ \cite{salovich}). We introduce a small repulsion $\lambda_{12}=-0.005$, which induces a phase shift between the two intraband pair amplitudes, $\varphi_{1}-\varphi_{2}=\pi$. In such a case, the coupling of the crosspairing pair amplitude with the intraband pair amplitudes (for $\lambda_{i3}>0$) will introduce frustration on the phase of the crosspairing order parameter $\varphi_{3}$. Phase frustration of similar sort is known in three-band systems \cite{stanev, n_orlova, trsb_3band} and can lead to skyrmionic vortex states \cite{3b_babaev, skyrm_garaud, skyr_orlova}, but is not possible in a two-band system unless crosspairing is present. In the present case, we reveal additional new physics, as crosspairing induces $s^{+-}\rightarrow s^{++}$ transition as a function of temperature, as shown in Fig. \ref{fig5}(a,b) for exemplified parameters of (BaK)Fe$_2$As$_2$. 

In the example shown in Fig. \ref{fig5}(a,b), after the transition, the pair amplitudes recover the same phase ($\theta=0$) until the expected BCS critical temperature of $\approx80$ K. In experiment however \cite{ding1}, the measured gaps abruptly cease at $T_c\approx40$ K, for reasons that are not understood to date. Without claiming to rigorously describe the non BCS behavior of the gaps versus temperature, we notice that our calculation of the gaps vs. temperature can closely reproduce the experimentally measured data [as shown in Fig. \ref{fig5}(c)], in cases that the $s^{+-}$ orbital antiphase is protected by symmetry or the transition to $s^{++}$ state is disallowed in any way.

In summary, although mostly neglected to date, the cross-band pairing in multiband superconductors is certainly of importance in materials with hybridized or energetically close bands in the vicinity of the Fermi level. In this regime, the interplay between intra- and cross-band pairing leads to several unique effects. For one, crossband pairing increases the splitting between intraband gaps, with a tendency to decrease the weaker gap towards an entirely novel ``gapless'' state, signatures of which will still be observable since vanishing gap does not imply vanishing order parameter(s) in this regime. The crosspairing also introduces the possibility of a phase frustration between the pairing channels, leading to novel transitions as a function of temperature (such as $s^{+-}\rightarrow s^{++}$), and likely nontrivial response of the superconductor to e.g., magnetic field \cite{change_vortex_core}. Our results call for revisiting the existing theories and experimental data for multiband superconductors with close bands, bearing also in mind that the band dispersions and chemical potential can be tuned towards a parameter regime where the above mentioned signatures of crosspairing can be detected. In that context, we point out the most recent measurements of Ref. \onlinecite{bergeal}, where tunability of multiple gaps has been achieved at the oxides' interface by gate doping around a Lifshitz transition, as the closest experimental system to our present model.

Besides the needed generalization to the case of multiple (3+) bands, the outlook of the present study is very broad. It includes understanding the effects of impurities, particularly magnetic ones where DOS signatures of crosspairing near a gapless state can overlap with the Majorana zero-energy bound state \cite{wang, yin2015}. It is also of interest to further examine the intra- to cross-pairing competition in the presence of spin-flip scattering \cite{gonelli}, oddness in parity \cite{odd_freq1}, and photo-induced phenomena \cite{kumar,porta}. Even beyond superconductivity, crosspairing and its competition with intraband pairing remains insufficiently explored in molecular optics \cite{coh_mol}, multicomponent superfluidity \cite{wilczek1}, and quantum chromodynamics \cite{wilczek2}.

The authors thank Andrea Guidini for his help during the initial stage of this work and Laura Fanfarillo for useful discussions. This work was partially supported by the Italian MIUR through the PRIN 2015 program (contract No. 2015C5SEJJ001) and the Research Foundation - Flanders (FWO). A. A. Vargas-Paredes acknowledges support by the joint doctoral program ``MultiSuper" (http://www.multisuper.org) and by the Erasmus+ exchange and M.V. Milo\v{s}evi\'c acknowledges the University of Camerino for support during his visit.

\nocite{*}

\bibliography{cross_biblio}

\providecommand{\noopsort}[1]{}\providecommand{\singleletter}[1]{#1}%
\begin{thebibliography}{63}%
\makeatletter
\providecommand \@ifxundefined [1]{%
 \@ifx{#1\undefined}
}%
\providecommand \@ifnum [1]{%
 \ifnum #1\expandafter \@firstoftwo
 \else \expandafter \@secondoftwo
 \fi
}%
\providecommand \@ifx [1]{%
 \ifx #1\expandafter \@firstoftwo
 \else \expandafter \@secondoftwo
 \fi
}%
\providecommand \natexlab [1]{#1}%
\providecommand \enquote  [1]{``#1''}%
\providecommand \bibnamefont  [1]{#1}%
\providecommand \bibfnamefont [1]{#1}%
\providecommand \citenamefont [1]{#1}%
\providecommand \href@noop [0]{\@secondoftwo}%
\providecommand \href [0]{\begingroup \@sanitize@url \@href}%
\providecommand \@href[1]{\@@startlink{#1}\@@href}%
\providecommand \@@href[1]{\endgroup#1\@@endlink}%
\providecommand \@sanitize@url [0]{\catcode `\\12\catcode `\$12\catcode
  `\&12\catcode `\#12\catcode `\^12\catcode `\_12\catcode `\%12\relax}%
\providecommand \@@startlink[1]{}%
\providecommand \@@endlink[0]{}%
\providecommand \url  [0]{\begingroup\@sanitize@url \@url }%
\providecommand \@url [1]{\endgroup\@href {#1}{\urlprefix }}%
\providecommand \urlprefix  [0]{URL }%
\providecommand \Eprint [0]{\href }%
\providecommand \doibase [0]{http://dx.doi.org/}%
\providecommand \selectlanguage [0]{\@gobble}%
\providecommand \bibinfo  [0]{\@secondoftwo}%
\providecommand \bibfield  [0]{\@secondoftwo}%
\providecommand \translation [1]{[#1]}%
\providecommand \BibitemOpen [0]{}%
\providecommand \bibitemStop [0]{}%
\providecommand \bibitemNoStop [0]{.\EOS\space}%
\providecommand \EOS [0]{\spacefactor3000\relax}%
\providecommand \BibitemShut  [1]{\csname bibitem#1\endcsname}%
\let\auto@bib@innerbib\@empty
\bibitem [{\citenamefont {Milo{\v{s}}evi{\'{c}}}\ and\ \citenamefont
  {Perali}(2015)}]{milo_perali}%
  \BibitemOpen
  \bibfield  {author} {\bibinfo {author} {\bibfnamefont {M.~V.}\ \bibnamefont
  {Milo{\v{s}}evi{\'{c}}}}\ and\ \bibinfo {author} {\bibfnamefont
  {A.}~\bibnamefont {Perali}},\ }\bibfield  {title} {\enquote {\bibinfo {title}
  {Emergent phenomena in multicomponent superconductivity: an introduction to
  the focus issue},}\ }\href {\doibase 10.1088/0953-2048/28/6/060201}
  {\bibfield  {journal} {\bibinfo  {journal} {Supercond. Sci. Technol.}\
  }\textbf {\bibinfo {volume} {28}},\ \bibinfo {pages} {060201} (\bibinfo
  {year} {2015})}\BibitemShut {NoStop}%
\bibitem [{\citenamefont {Giorgianni}\ \emph {et~al.}(2019)\citenamefont
  {Giorgianni}, \citenamefont {Cea}, \citenamefont {Vicario}, \citenamefont
  {Hauri}, \citenamefont {Withanage}, \citenamefont {Xi},\ and\ \citenamefont
  {Benfatto}}]{giorgianni}%
  \BibitemOpen
  \bibfield  {author} {\bibinfo {author} {\bibfnamefont {F.}~\bibnamefont
  {Giorgianni}}, \bibinfo {author} {\bibfnamefont {T.}~\bibnamefont {Cea}},
  \bibinfo {author} {\bibfnamefont {C.}~\bibnamefont {Vicario}}, \bibinfo
  {author} {\bibfnamefont {C.~P.}\ \bibnamefont {Hauri}}, \bibinfo {author}
  {\bibfnamefont {W.~K.}\ \bibnamefont {Withanage}}, \bibinfo {author}
  {\bibfnamefont {X.}~\bibnamefont {Xi}}, \ and\ \bibinfo {author}
  {\bibfnamefont {L.}~\bibnamefont {Benfatto}},\ }\bibfield  {title} {\enquote
  {\bibinfo {title} {{Leggett mode controlled by light pulses}},}\ }\href
  {\doibase 10.1038/s41567-018-0385-4} {\bibfield  {journal} {\bibinfo
  {journal} {Nature Physics}\ }\textbf {\bibinfo {volume} {15}},\ \bibinfo
  {pages} {341--346} (\bibinfo {year} {2019})}\BibitemShut {NoStop}%
\bibitem [{\citenamefont {Hanaguri}\ \emph {et~al.}(2019)\citenamefont
  {Hanaguri}, \citenamefont {Kasahara}, \citenamefont {B\"oker}, \citenamefont
  {Eremin}, \citenamefont {Shibauchi},\ and\ \citenamefont
  {Matsuda}}]{hanaguri}%
  \BibitemOpen
  \bibfield  {author} {\bibinfo {author} {\bibfnamefont {T.}~\bibnamefont
  {Hanaguri}}, \bibinfo {author} {\bibfnamefont {S.}~\bibnamefont {Kasahara}},
  \bibinfo {author} {\bibfnamefont {J.}~\bibnamefont {B\"oker}}, \bibinfo
  {author} {\bibfnamefont {I.}~\bibnamefont {Eremin}}, \bibinfo {author}
  {\bibfnamefont {T.}~\bibnamefont {Shibauchi}}, \ and\ \bibinfo {author}
  {\bibfnamefont {Y.}~\bibnamefont {Matsuda}},\ }\bibfield  {title} {\enquote
  {\bibinfo {title} {{Quantum Vortex Core and Missing Pseudogap in the
  Multiband BCS-BEC Crossover Superconductor FeSe}},}\ }\href {\doibase
  10.1103/PhysRevLett.122.077001} {\bibfield  {journal} {\bibinfo  {journal}
  {Phys. Rev. Lett.}\ }\textbf {\bibinfo {volume} {122}},\ \bibinfo {pages}
  {077001} (\bibinfo {year} {2019})}\BibitemShut {NoStop}%
\bibitem [{\citenamefont {Singh}\ \emph {et~al.}(2019)\citenamefont {Singh},
  \citenamefont {Jouan}, \citenamefont {Herranz}, \citenamefont {Scigaj},
  \citenamefont {S\'anchez}, \citenamefont {Benfatto}, \citenamefont {Caprara},
  \citenamefont {Grilli}, \citenamefont {Saiz}, \citenamefont {Cou\"edo},
  \citenamefont {Feuillet-Palma}, \citenamefont {Lesueur},\ and\ \citenamefont
  {Bergeal}}]{bergeal}%
  \BibitemOpen
  \bibfield  {author} {\bibinfo {author} {\bibfnamefont {G.}~\bibnamefont
  {Singh}}, \bibinfo {author} {\bibfnamefont {A.}~\bibnamefont {Jouan}},
  \bibinfo {author} {\bibfnamefont {G.}~\bibnamefont {Herranz}}, \bibinfo
  {author} {\bibfnamefont {M.}~\bibnamefont {Scigaj}}, \bibinfo {author}
  {\bibfnamefont {F.}~\bibnamefont {S\'anchez}}, \bibinfo {author}
  {\bibfnamefont {L.}~\bibnamefont {Benfatto}}, \bibinfo {author}
  {\bibfnamefont {S.}~\bibnamefont {Caprara}}, \bibinfo {author} {\bibfnamefont
  {M.}~\bibnamefont {Grilli}}, \bibinfo {author} {\bibfnamefont
  {G.}~\bibnamefont {Saiz}}, \bibinfo {author} {\bibfnamefont {F.}~\bibnamefont
  {Cou\"edo}}, \bibinfo {author} {\bibfnamefont {C.}~\bibnamefont
  {Feuillet-Palma}}, \bibinfo {author} {\bibfnamefont {J.}~\bibnamefont
  {Lesueur}}, \ and\ \bibinfo {author} {\bibfnamefont {N.}~\bibnamefont
  {Bergeal}},\ }\bibfield  {title} {\enquote {\bibinfo {title} {Gap suppression
  at a lifshitz transition in a multi-condensate superconductor},}\ }\href
  {\doibase 10.1038/s41563-019-0354-z} {\bibfield  {journal} {\bibinfo
  {journal} {Nature Materials}\ } (\bibinfo {year} {2019}),\
  10.1038/s41563-019-0354-z}\BibitemShut {NoStop}%
\bibitem [{\citenamefont {Li}\ \emph {et~al.}(2018)\citenamefont {Li},
  \citenamefont {An}, \citenamefont {Hua}, \citenamefont {Chen}, \citenamefont
  {Zhou}, \citenamefont {Zhou}, \citenamefont {Zhang}, \citenamefont {Park},
  \citenamefont {Wang}, \citenamefont {Lu}, \citenamefont {Zheng},
  \citenamefont {Yang},\ and\ \citenamefont {Xu}}]{li}%
  \BibitemOpen
  \bibfield  {author} {\bibinfo {author} {\bibfnamefont {Y.}~\bibnamefont
  {Li}}, \bibinfo {author} {\bibfnamefont {C.}~\bibnamefont {An}}, \bibinfo
  {author} {\bibfnamefont {C.}~\bibnamefont {Hua}}, \bibinfo {author}
  {\bibfnamefont {X.}~\bibnamefont {Chen}}, \bibinfo {author} {\bibfnamefont
  {Y.}~\bibnamefont {Zhou}}, \bibinfo {author} {\bibfnamefont {Y.}~\bibnamefont
  {Zhou}}, \bibinfo {author} {\bibfnamefont {R.}~\bibnamefont {Zhang}},
  \bibinfo {author} {\bibfnamefont {C.}~\bibnamefont {Park}}, \bibinfo {author}
  {\bibfnamefont {Z.}~\bibnamefont {Wang}}, \bibinfo {author} {\bibfnamefont
  {Y.}~\bibnamefont {Lu}}, \bibinfo {author} {\bibfnamefont {Y.}~\bibnamefont
  {Zheng}}, \bibinfo {author} {\bibfnamefont {Z.}~\bibnamefont {Yang}}, \ and\
  \bibinfo {author} {\bibfnamefont {Z.}~\bibnamefont {Xu}},\ }\bibfield
  {title} {\enquote {\bibinfo {title} {{Pressure-induced superconductivity in
  topological semimetal NbAs2}},}\ }\href {\doibase 10.1038/s41535-018-0132-1}
  {\bibfield  {journal} {\bibinfo  {journal} {npj Quantum Materials}\ }\textbf
  {\bibinfo {volume} {3}},\ \bibinfo {pages} {58} (\bibinfo {year}
  {2018})}\BibitemShut {NoStop}%
\bibitem [{\citenamefont {Li}\ \emph {et~al.}(2010)\citenamefont {Li},
  \citenamefont {Zeng}, \citenamefont {Lu},\ and\ \citenamefont {Dou}}]{wx_li}%
  \BibitemOpen
  \bibfield  {author} {\bibinfo {author} {\bibfnamefont {W.X.}\ \bibnamefont
  {Li}}, \bibinfo {author} {\bibfnamefont {R.}~\bibnamefont {Zeng}}, \bibinfo
  {author} {\bibfnamefont {L.}~\bibnamefont {Lu}}, \ and\ \bibinfo {author}
  {\bibfnamefont {S.X.}\ \bibnamefont {Dou}},\ }\bibfield  {title} {\enquote
  {\bibinfo {title} {{Dependence of superconducting properties on lattice
  strain in MgB$_2$}},}\ }\href {\doibase 10.1016/j.physc.2009.11.053}
  {\bibfield  {journal} {\bibinfo  {journal} {Physica C: Superconductivity and
  its Applications}\ }\textbf {\bibinfo {volume} {470}},\ \bibinfo {pages}
  {S629 -- S630} (\bibinfo {year} {2010})},\ \bibinfo {note} {proceedings of
  the 9th International Conference on Materials and Mechanisms of
  Superconductivity}\BibitemShut {NoStop}%
\bibitem [{\citenamefont {Costanzo}\ \emph {et~al.}(2016)\citenamefont
  {Costanzo}, \citenamefont {Jo}, \citenamefont {Berger},\ and\ \citenamefont
  {Morpurgo}}]{costanzo}%
  \BibitemOpen
  \bibfield  {author} {\bibinfo {author} {\bibfnamefont {D.}~\bibnamefont
  {Costanzo}}, \bibinfo {author} {\bibfnamefont {S.}~\bibnamefont {Jo}},
  \bibinfo {author} {\bibfnamefont {H.}~\bibnamefont {Berger}}, \ and\ \bibinfo
  {author} {\bibfnamefont {A.~F.}\ \bibnamefont {Morpurgo}},\ }\bibfield
  {title} {\enquote {\bibinfo {title} {{Gate-induced superconductivity in
  atomically thin MoS2 crystals}},}\ }\href
  {https://doi.org/10.1038/nnano.2015.314 http://10.0.4.14/nnano.2015.314
  https://www.nature.com/articles/nnano.2015.314{\#}supplementary-information}
  {\bibfield  {journal} {\bibinfo  {journal} {Nature Nanotechnology}\ }\textbf
  {\bibinfo {volume} {11}},\ \bibinfo {pages} {339} (\bibinfo {year}
  {2016})}\BibitemShut {NoStop}%
\bibitem [{\citenamefont {Continenza}\ and\ \citenamefont
  {Profeta}(2018)}]{continenza}%
  \BibitemOpen
  \bibfield  {author} {\bibinfo {author} {\bibfnamefont {A.}~\bibnamefont
  {Continenza}}\ and\ \bibinfo {author} {\bibfnamefont {G.}~\bibnamefont
  {Profeta}},\ }\bibfield  {title} {\enquote {\bibinfo {title} {{Chemical
  doping in pnictides superconductors: The case of
  Ca(Fe$_{1-x}$X$_x$)$_2$As$_2$, X $=$ Co, Ni, Pt}},}\ }\href {\doibase
  10.1016/j.jmmm.2017.12.060} {\bibfield  {journal} {\bibinfo  {journal}
  {Journal of Magnetism and Magnetic Materials}\ }\textbf {\bibinfo {volume}
  {452}},\ \bibinfo {pages} {179 -- 183} (\bibinfo {year} {2018})}\BibitemShut
  {NoStop}%
\bibitem [{\citenamefont {Porta}\ \emph {et~al.}()\citenamefont {Porta},
  \citenamefont {Privitera}, \citenamefont {Traverso~Ziani}, \citenamefont
  {Sassetti}, \citenamefont {Cavaliere},\ and\ \citenamefont
  {Trauzettel}}]{porta}%
  \BibitemOpen
  \bibfield  {author} {\bibinfo {author} {\bibfnamefont {S.}~\bibnamefont
  {Porta}}, \bibinfo {author} {\bibfnamefont {L.}~\bibnamefont {Privitera}},
  \bibinfo {author} {\bibfnamefont {N.}~\bibnamefont {Traverso~Ziani}},
  \bibinfo {author} {\bibfnamefont {M.}~\bibnamefont {Sassetti}}, \bibinfo
  {author} {\bibfnamefont {F.}~\bibnamefont {Cavaliere}}, \ and\ \bibinfo
  {author} {\bibfnamefont {B.}~\bibnamefont {Trauzettel}},\ }\bibfield  {title}
  {\enquote {\bibinfo {title} {{Feasible model for photo-induced interband
  pairing}},}\ }\href@noop {} {\ }\Eprint {http://arxiv.org/abs/1903.12396}
  {arXiv:1903.12396 [cond-mat.mes-hall]} \BibitemShut {NoStop}%
\bibitem [{\citenamefont {Nagamatsu}\ \emph {et~al.}(2001)\citenamefont
  {Nagamatsu}, \citenamefont {Nagakawa}, \citenamefont {Muranaka},
  \citenamefont {Zenitani},\ and\ \citenamefont {Akimitsu}}]{akimitsu}%
  \BibitemOpen
  \bibfield  {author} {\bibinfo {author} {\bibfnamefont {J.}~\bibnamefont
  {Nagamatsu}}, \bibinfo {author} {\bibfnamefont {N.}~\bibnamefont {Nagakawa}},
  \bibinfo {author} {\bibfnamefont {T.}~\bibnamefont {Muranaka}}, \bibinfo
  {author} {\bibfnamefont {Y.}~\bibnamefont {Zenitani}}, \ and\ \bibinfo
  {author} {\bibfnamefont {J.}~\bibnamefont {Akimitsu}},\ }\bibfield  {title}
  {\enquote {\bibinfo {title} {Superconductivity at 39 k in magnesium
  diboride},}\ }\href {https://doi.org/10.1038/nature01619
  http://10.0.4.14/nature01619} {\bibfield  {journal} {\bibinfo  {journal}
  {Nature}\ }\textbf {\bibinfo {volume} {410}},\ \bibinfo {pages} {63}
  (\bibinfo {year} {2001})}\BibitemShut {NoStop}%
\bibitem [{\citenamefont {de' Medici}\ \emph {et~al.}(2014)\citenamefont {de'
  Medici}, \citenamefont {Giovannetti},\ and\ \citenamefont
  {Capone}}]{selective_mott}%
  \BibitemOpen
  \bibfield  {author} {\bibinfo {author} {\bibfnamefont {L.}~\bibnamefont {de'
  Medici}}, \bibinfo {author} {\bibfnamefont {G.}~\bibnamefont {Giovannetti}},
  \ and\ \bibinfo {author} {\bibfnamefont {M.}~\bibnamefont {Capone}},\
  }\bibfield  {title} {\enquote {\bibinfo {title} {{Selective Mott Physics as a
  Key to Iron Superconductors}},}\ }\href {\doibase
  10.1103/PhysRevLett.112.177001} {\bibfield  {journal} {\bibinfo  {journal}
  {Phys. Rev. Lett.}\ }\textbf {\bibinfo {volume} {112}},\ \bibinfo {pages}
  {177001} (\bibinfo {year} {2014})}\BibitemShut {NoStop}%
\bibitem [{\citenamefont {Benfatto}\ \emph {et~al.}(2018)\citenamefont
  {Benfatto}, \citenamefont {Valenzuela},\ and\ \citenamefont
  {Fanfarillo}}]{nemat_pairing_orbital_select}%
  \BibitemOpen
  \bibfield  {author} {\bibinfo {author} {\bibfnamefont {L.}~\bibnamefont
  {Benfatto}}, \bibinfo {author} {\bibfnamefont {B.}~\bibnamefont
  {Valenzuela}}, \ and\ \bibinfo {author} {\bibfnamefont {L.}~\bibnamefont
  {Fanfarillo}},\ }\bibfield  {title} {\enquote {\bibinfo {title} {{Nematic
  pairing from orbital-selective spin fluctuations in FeSe}},}\ }\href
  {\doibase 10.1038/s41535-018-0129-9} {\bibfield  {journal} {\bibinfo
  {journal} {npj Quantum Materials}\ }\textbf {\bibinfo {volume} {3}},\
  \bibinfo {pages} {56} (\bibinfo {year} {2018})}\BibitemShut {NoStop}%
\bibitem [{\citenamefont {Yu}\ \emph {et~al.}(2018)\citenamefont {Yu},
  \citenamefont {Zhu},\ and\ \citenamefont {Si}}]{orbital_selec1}%
  \BibitemOpen
  \bibfield  {author} {\bibinfo {author} {\bibfnamefont {R.}~\bibnamefont
  {Yu}}, \bibinfo {author} {\bibfnamefont {J.}~\bibnamefont {Zhu}}, \ and\
  \bibinfo {author} {\bibfnamefont {Q.}~\bibnamefont {Si}},\ }\bibfield
  {title} {\enquote {\bibinfo {title} {{Orbital Selectivity Enhanced by Nematic
  Order in FeSe}},}\ }\href {\doibase 10.1103/PhysRevLett.121.227003}
  {\bibfield  {journal} {\bibinfo  {journal} {Phys. Rev. Lett.}\ }\textbf
  {\bibinfo {volume} {121}},\ \bibinfo {pages} {227003} (\bibinfo {year}
  {2018})}\BibitemShut {NoStop}%
\bibitem [{\citenamefont {Herbrych}\ \emph {et~al.}(2018)\citenamefont
  {Herbrych}, \citenamefont {Kaushal}, \citenamefont {Nocera}, \citenamefont
  {Alvarez}, \citenamefont {Moreo},\ and\ \citenamefont
  {Dagotto}}]{orbital_selec2}%
  \BibitemOpen
  \bibfield  {author} {\bibinfo {author} {\bibfnamefont {J.}~\bibnamefont
  {Herbrych}}, \bibinfo {author} {\bibfnamefont {N.}~\bibnamefont {Kaushal}},
  \bibinfo {author} {\bibfnamefont {A.}~\bibnamefont {Nocera}}, \bibinfo
  {author} {\bibfnamefont {G.}~\bibnamefont {Alvarez}}, \bibinfo {author}
  {\bibfnamefont {A.}~\bibnamefont {Moreo}}, \ and\ \bibinfo {author}
  {\bibfnamefont {E.}~\bibnamefont {Dagotto}},\ }\bibfield  {title} {\enquote
  {\bibinfo {title} {{Spin dynamics of the block orbital-selective Mott
  phase}},}\ }\href {\doibase 10.1038/s41467-018-06181-6} {\bibfield  {journal}
  {\bibinfo  {journal} {Nature Communications}\ }\textbf {\bibinfo {volume}
  {9}},\ \bibinfo {pages} {3736} (\bibinfo {year} {2018})}\BibitemShut
  {NoStop}%
\bibitem [{\citenamefont {Black-Schaffer}\ and\ \citenamefont
  {Balatsky}(2013)}]{odd_freq1}%
  \BibitemOpen
  \bibfield  {author} {\bibinfo {author} {\bibfnamefont {A.~M.}\ \bibnamefont
  {Black-Schaffer}}\ and\ \bibinfo {author} {\bibfnamefont {A.~V.}\
  \bibnamefont {Balatsky}},\ }\bibfield  {title} {\enquote {\bibinfo {title}
  {Odd-frequency superconducting pairing in multiband superconductors},}\
  }\href {\doibase 10.1103/PhysRevB.88.104514} {\bibfield  {journal} {\bibinfo
  {journal} {Phys. Rev. B}\ }\textbf {\bibinfo {volume} {88}},\ \bibinfo
  {pages} {104514} (\bibinfo {year} {2013})}\BibitemShut {NoStop}%
\bibitem [{\citenamefont {Ding}\ \emph {et~al.}(2008)\citenamefont {Ding},
  \citenamefont {Richard}, \citenamefont {Nakayama}, \citenamefont {Sugawara},
  \citenamefont {Arakane}, \citenamefont {Sekiba}, \citenamefont {Takayama},
  \citenamefont {Souma}, \citenamefont {Sato}, \citenamefont {Takahashi},
  \citenamefont {Wang}, \citenamefont {Dai}, \citenamefont {Fang},
  \citenamefont {Chen}, \citenamefont {Luo},\ and\ \citenamefont
  {Wang}}]{ding1}%
  \BibitemOpen
  \bibfield  {author} {\bibinfo {author} {\bibfnamefont {H.}~\bibnamefont
  {Ding}}, \bibinfo {author} {\bibfnamefont {P.}~\bibnamefont {Richard}},
  \bibinfo {author} {\bibfnamefont {K.}~\bibnamefont {Nakayama}}, \bibinfo
  {author} {\bibfnamefont {K.}~\bibnamefont {Sugawara}}, \bibinfo {author}
  {\bibfnamefont {T.}~\bibnamefont {Arakane}}, \bibinfo {author} {\bibfnamefont
  {Y.}~\bibnamefont {Sekiba}}, \bibinfo {author} {\bibfnamefont
  {A.}~\bibnamefont {Takayama}}, \bibinfo {author} {\bibfnamefont
  {S.}~\bibnamefont {Souma}}, \bibinfo {author} {\bibfnamefont
  {T.}~\bibnamefont {Sato}}, \bibinfo {author} {\bibfnamefont {T.}~\bibnamefont
  {Takahashi}}, \bibinfo {author} {\bibfnamefont {Z.}~\bibnamefont {Wang}},
  \bibinfo {author} {\bibfnamefont {X.}~\bibnamefont {Dai}}, \bibinfo {author}
  {\bibfnamefont {Z.}~\bibnamefont {Fang}}, \bibinfo {author} {\bibfnamefont
  {G.~F.}\ \bibnamefont {Chen}}, \bibinfo {author} {\bibfnamefont {J.~L.}\
  \bibnamefont {Luo}}, \ and\ \bibinfo {author} {\bibfnamefont {N.~L.}\
  \bibnamefont {Wang}},\ }\bibfield  {title} {\enquote {\bibinfo {title}
  {{Observation of Fermi-surface{\textendash}dependent nodeless superconducting
  gaps in Ba$_{0.6}$K$_{0.4}$Fe$_2$As$_2$}},}\ }\href {\doibase
  10.1209/0295-5075/83/47001} {\bibfield  {journal} {\bibinfo  {journal}
  {{EPL}}\ }\textbf {\bibinfo {volume} {83}},\ \bibinfo {pages} {47001}
  (\bibinfo {year} {2008})}\BibitemShut {NoStop}%
\bibitem [{\citenamefont {Bianconi}\ \emph {et~al.}(1997)\citenamefont
  {Bianconi}, \citenamefont {Valletta}, \citenamefont {Perali},\ and\
  \citenamefont {Saini}}]{bianconi97}%
  \BibitemOpen
  \bibfield  {author} {\bibinfo {author} {\bibfnamefont {A.}~\bibnamefont
  {Bianconi}}, \bibinfo {author} {\bibfnamefont {A.}~\bibnamefont {Valletta}},
  \bibinfo {author} {\bibfnamefont {A.}~\bibnamefont {Perali}}, \ and\ \bibinfo
  {author} {\bibfnamefont {N.L.}\ \bibnamefont {Saini}},\ }\bibfield  {title}
  {\enquote {\bibinfo {title} {High tc superconductivity in a superlattice of
  quantum stripes},}\ }\href {\doibase 10.1016/S0038-1098(97)00011-2}
  {\bibfield  {journal} {\bibinfo  {journal} {Solid State Communications}\
  }\textbf {\bibinfo {volume} {102}},\ \bibinfo {pages} {369 -- 374} (\bibinfo
  {year} {1997})}\BibitemShut {NoStop}%
\bibitem [{\citenamefont {Valletta}\ \emph {et~al.}(1997)\citenamefont
  {Valletta}, \citenamefont {Bianconi}, \citenamefont {Perali},\ and\
  \citenamefont {Saini}}]{valletta97}%
  \BibitemOpen
  \bibfield  {author} {\bibinfo {author} {\bibfnamefont {A.}~\bibnamefont
  {Valletta}}, \bibinfo {author} {\bibfnamefont {A.}~\bibnamefont {Bianconi}},
  \bibinfo {author} {\bibfnamefont {A.}~\bibnamefont {Perali}}, \ and\ \bibinfo
  {author} {\bibfnamefont {N.L.}\ \bibnamefont {Saini}},\ }\bibfield  {title}
  {\enquote {\bibinfo {title} {Electronic and superconducting properties of a
  superlattice of quantum stripes at the atomic limit},}\ }\href {\doibase
  10.1007/s002570050513} {\bibfield  {journal} {\bibinfo  {journal}
  {Zeitschrift f{\"u}r Physik B Condensed Matter}\ }\textbf {\bibinfo {volume}
  {104}},\ \bibinfo {pages} {707--713} (\bibinfo {year} {1997})}\BibitemShut
  {NoStop}%
\bibitem [{\citenamefont {Shanenko}\ \emph {et~al.}(2015)\citenamefont
  {Shanenko}, \citenamefont {Albino~Aguiar}, \citenamefont {Vagov},
  \citenamefont {Croitoru},\ and\ \citenamefont
  {Milo\u{s}evi\'c}}]{cross_arkady}%
  \BibitemOpen
  \bibfield  {author} {\bibinfo {author} {\bibfnamefont {A.~A.}\ \bibnamefont
  {Shanenko}}, \bibinfo {author} {\bibfnamefont {J.}~\bibnamefont
  {Albino~Aguiar}}, \bibinfo {author} {\bibfnamefont {A.}~\bibnamefont
  {Vagov}}, \bibinfo {author} {\bibfnamefont {M.~D.}\ \bibnamefont {Croitoru}},
  \ and\ \bibinfo {author} {\bibfnamefont {M.~V.}\ \bibnamefont
  {Milo\u{s}evi\'c}},\ }\bibfield  {title} {\enquote {\bibinfo {title}
  {Atomically flat superconducting nanofilms: multiband properties and
  mean-field theory},}\ }\href
  {http://stacks.iop.org/0953-2048/28/i=5/a=054001} {\bibfield  {journal}
  {\bibinfo  {journal} {Supercond. Sci. Technol.}\ }\textbf {\bibinfo {volume}
  {28}},\ \bibinfo {pages} {054001} (\bibinfo {year} {2015})}\BibitemShut
  {NoStop}%
\bibitem [{\citenamefont {Flammia}\ \emph {et~al.}(2018)\citenamefont
  {Flammia}, \citenamefont {Zhang}, \citenamefont {Covaci}, \citenamefont
  {Perali},\ and\ \citenamefont {Milo\v{s}evi\'c}}]{sc_nanoribbon}%
  \BibitemOpen
  \bibfield  {author} {\bibinfo {author} {\bibfnamefont {L.}~\bibnamefont
  {Flammia}}, \bibinfo {author} {\bibfnamefont {L.-F.}\ \bibnamefont {Zhang}},
  \bibinfo {author} {\bibfnamefont {L.}~\bibnamefont {Covaci}}, \bibinfo
  {author} {\bibfnamefont {A.}~\bibnamefont {Perali}}, \ and\ \bibinfo {author}
  {\bibfnamefont {M.~V.}\ \bibnamefont {Milo\v{s}evi\'c}},\ }\bibfield  {title}
  {\enquote {\bibinfo {title} {Superconducting nanoribbon with a constriction:
  A quantum-confined josephson junction},}\ }\href {\doibase
  10.1103/PhysRevB.97.134514} {\bibfield  {journal} {\bibinfo  {journal} {Phys.
  Rev. B}\ }\textbf {\bibinfo {volume} {97}},\ \bibinfo {pages} {134514}
  (\bibinfo {year} {2018})}\BibitemShut {NoStop}%
\bibitem [{\citenamefont {Zhang}\ \emph
  {et~al.}(2017{\natexlab{a}})\citenamefont {Zhang}, \citenamefont {Flammia},
  \citenamefont {Covaci}, \citenamefont {Perali},\ and\ \citenamefont
  {Milo\v{s}evi\'c}}]{surface_step}%
  \BibitemOpen
  \bibfield  {author} {\bibinfo {author} {\bibfnamefont {L.-F.}\ \bibnamefont
  {Zhang}}, \bibinfo {author} {\bibfnamefont {L.}~\bibnamefont {Flammia}},
  \bibinfo {author} {\bibfnamefont {L.}~\bibnamefont {Covaci}}, \bibinfo
  {author} {\bibfnamefont {A.}~\bibnamefont {Perali}}, \ and\ \bibinfo {author}
  {\bibfnamefont {M.~V.}\ \bibnamefont {Milo\v{s}evi\'c}},\ }\bibfield  {title}
  {\enquote {\bibinfo {title} {Multifaceted impact of a surface step on
  superconductivity in atomically thin films},}\ }\href {\doibase
  10.1103/PhysRevB.96.104509} {\bibfield  {journal} {\bibinfo  {journal} {Phys.
  Rev. B}\ }\textbf {\bibinfo {volume} {96}},\ \bibinfo {pages} {104509}
  (\bibinfo {year} {2017}{\natexlab{a}})}\BibitemShut {NoStop}%
\bibitem [{\citenamefont {Valentinis}\ \emph {et~al.}(2017)\citenamefont
  {Valentinis}, \citenamefont {Gariglio}, \citenamefont {F\^ete}, \citenamefont
  {Triscone}, \citenamefont {Berthod},\ and\ \citenamefont {van~der
  Marel}}]{valentinis}%
  \BibitemOpen
  \bibfield  {author} {\bibinfo {author} {\bibfnamefont {D.}~\bibnamefont
  {Valentinis}}, \bibinfo {author} {\bibfnamefont {S.}~\bibnamefont
  {Gariglio}}, \bibinfo {author} {\bibfnamefont {A.}~\bibnamefont {F\^ete}},
  \bibinfo {author} {\bibfnamefont {J.-M.}\ \bibnamefont {Triscone}}, \bibinfo
  {author} {\bibfnamefont {C.}~\bibnamefont {Berthod}}, \ and\ \bibinfo
  {author} {\bibfnamefont {D.}~\bibnamefont {van~der Marel}},\ }\bibfield
  {title} {\enquote {\bibinfo {title} {{Modulation of the superconducting
  critical temperature due to quantum confinement at the LaAlO$_3/$SrTiO$_3$
  interface}},}\ }\href {\doibase 10.1103/PhysRevB.96.094518} {\bibfield
  {journal} {\bibinfo  {journal} {Phys. Rev. B}\ }\textbf {\bibinfo {volume}
  {96}},\ \bibinfo {pages} {094518} (\bibinfo {year} {2017})}\BibitemShut
  {NoStop}%
\bibitem [{\citenamefont {Mohanta}\ and\ \citenamefont
  {Taraphder}(2015)}]{mohanta}%
  \BibitemOpen
  \bibfield  {author} {\bibinfo {author} {\bibfnamefont {N.}~\bibnamefont
  {Mohanta}}\ and\ \bibinfo {author} {\bibfnamefont {A.}~\bibnamefont
  {Taraphder}},\ }\bibfield  {title} {\enquote {\bibinfo {title} {{Multiband
  theory of superconductivity at the
  ${\mathrm{LaAlO}}_{3}/{\mathrm{SrTiO}}_{3}$ interface}},}\ }\href {\doibase
  10.1103/PhysRevB.92.174531} {\bibfield  {journal} {\bibinfo  {journal} {Phys.
  Rev. B}\ }\textbf {\bibinfo {volume} {92}},\ \bibinfo {pages} {174531}
  (\bibinfo {year} {2015})}\BibitemShut {NoStop}%
\bibitem [{\citenamefont {Trevisan}\ \emph {et~al.}(2018)\citenamefont
  {Trevisan}, \citenamefont {Sch\"utt},\ and\ \citenamefont
  {Fernandes}}]{thais1}%
  \BibitemOpen
  \bibfield  {author} {\bibinfo {author} {\bibfnamefont {T.~V.}\ \bibnamefont
  {Trevisan}}, \bibinfo {author} {\bibfnamefont {M.}~\bibnamefont {Sch\"utt}},
  \ and\ \bibinfo {author} {\bibfnamefont {R.~M.}\ \bibnamefont {Fernandes}},\
  }\bibfield  {title} {\enquote {\bibinfo {title} {{Unconventional Multiband
  Superconductivity in Bulk ${\mathrm{SrTiO}}_{3}$ and
  ${\mathrm{LaAlO}}_{3}/{\mathrm{SrTiO}}_{3}$ Interfaces}},}\ }\href {\doibase
  10.1103/PhysRevLett.121.127002} {\bibfield  {journal} {\bibinfo  {journal}
  {Phys. Rev. Lett.}\ }\textbf {\bibinfo {volume} {121}},\ \bibinfo {pages}
  {127002} (\bibinfo {year} {2018})}\BibitemShut {NoStop}%
\bibitem [{\citenamefont {Wang}\ \emph {et~al.}(2017)\citenamefont {Wang},
  \citenamefont {Gao}, \citenamefont {Huang},\ and\ \citenamefont
  {Chen}}]{chain_molecule}%
  \BibitemOpen
  \bibfield  {author} {\bibinfo {author} {\bibfnamefont {R.-S.}\ \bibnamefont
  {Wang}}, \bibinfo {author} {\bibfnamefont {Y.}~\bibnamefont {Gao}}, \bibinfo
  {author} {\bibfnamefont {Z.-B.}\ \bibnamefont {Huang}}, \ and\ \bibinfo
  {author} {\bibfnamefont {X.-J.}\ \bibnamefont {Chen}},\ }\bibfield  {title}
  {\enquote {\bibinfo {title} {Superconductivity above 120 kelvin in a chain
  link molecule},}\ }\href@noop {} {\bibfield  {journal} {\bibinfo  {journal}
  {arXiv e-prints}\ ,\ \bibinfo {eid} {arXiv:1703.06641}} (\bibinfo {year}
  {2017})},\ \Eprint {http://arxiv.org/abs/1703.06641} {arXiv:1703.06641
  [cond-mat.supr-con]} \BibitemShut {NoStop}%
\bibitem [{\citenamefont {Zhang}\ \emph
  {et~al.}(2017{\natexlab{b}})\citenamefont {Zhang}, \citenamefont {Zhou},
  \citenamefont {Cui}, \citenamefont {Zhao},\ and\ \citenamefont
  {Liu}}]{met_org}%
  \BibitemOpen
  \bibfield  {author} {\bibinfo {author} {\bibfnamefont {X.}~\bibnamefont
  {Zhang}}, \bibinfo {author} {\bibfnamefont {Y.}~\bibnamefont {Zhou}},
  \bibinfo {author} {\bibfnamefont {B.}~\bibnamefont {Cui}}, \bibinfo {author}
  {\bibfnamefont {M.}~\bibnamefont {Zhao}}, \ and\ \bibinfo {author}
  {\bibfnamefont {F.}~\bibnamefont {Liu}},\ }\bibfield  {title} {\enquote
  {\bibinfo {title} {{Theoretical Discovery of a Superconducting
  Two-Dimensional Metal-Organic Framework}},}\ }\href {\doibase
  10.1021/acs.nanolett.7b02795} {\bibfield  {journal} {\bibinfo  {journal}
  {Nano Letters}\ }\textbf {\bibinfo {volume} {17}},\ \bibinfo {pages}
  {6166--6170} (\bibinfo {year} {2017}{\natexlab{b}})}\BibitemShut {NoStop}%
\bibitem [{\citenamefont {Mazziotti}\ \emph {et~al.}(2017)\citenamefont
  {Mazziotti}, \citenamefont {Valletta}, \citenamefont {Campi}, \citenamefont
  {Innocenti}, \citenamefont {Perali},\ and\ \citenamefont
  {Bianconi}}]{possible_fano}%
  \BibitemOpen
  \bibfield  {author} {\bibinfo {author} {\bibfnamefont {M.~V.}\ \bibnamefont
  {Mazziotti}}, \bibinfo {author} {\bibfnamefont {A.}~\bibnamefont {Valletta}},
  \bibinfo {author} {\bibfnamefont {G.}~\bibnamefont {Campi}}, \bibinfo
  {author} {\bibfnamefont {D.}~\bibnamefont {Innocenti}}, \bibinfo {author}
  {\bibfnamefont {A.}~\bibnamefont {Perali}}, \ and\ \bibinfo {author}
  {\bibfnamefont {A.}~\bibnamefont {Bianconi}},\ }\bibfield  {title} {\enquote
  {\bibinfo {title} {Possible fano resonance for high-tc multi-gap
  superconductivity in p-terphenyl doped by k at the lifshitz transition},}\
  }\href {\doibase 10.1209/0295-5075/118/37003} {\bibfield  {journal} {\bibinfo
   {journal} {{EPL}}\ }\textbf {\bibinfo {volume} {118}},\ \bibinfo {pages}
  {37003} (\bibinfo {year} {2017})}\BibitemShut {NoStop}%
\bibitem [{\citenamefont {Liu}\ and\ \citenamefont {Wilczek}(2003)}]{wilczek1}%
  \BibitemOpen
  \bibfield  {author} {\bibinfo {author} {\bibfnamefont {W.~V.}\ \bibnamefont
  {Liu}}\ and\ \bibinfo {author} {\bibfnamefont {F.}~\bibnamefont {Wilczek}},\
  }\bibfield  {title} {\enquote {\bibinfo {title} {Interior gap
  superfluidity},}\ }\href {\doibase 10.1103/PhysRevLett.90.047002} {\bibfield
  {journal} {\bibinfo  {journal} {Phys. Rev. Lett.}\ }\textbf {\bibinfo
  {volume} {90}},\ \bibinfo {pages} {047002} (\bibinfo {year}
  {2003})}\BibitemShut {NoStop}%
\bibitem [{\citenamefont {Gubankova}\ \emph {et~al.}(2003)\citenamefont
  {Gubankova}, \citenamefont {Liu},\ and\ \citenamefont {Wilczek}}]{wilczek2}%
  \BibitemOpen
  \bibfield  {author} {\bibinfo {author} {\bibfnamefont {E.}~\bibnamefont
  {Gubankova}}, \bibinfo {author} {\bibfnamefont {W.~V.}\ \bibnamefont {Liu}},
  \ and\ \bibinfo {author} {\bibfnamefont {F.}~\bibnamefont {Wilczek}},\
  }\bibfield  {title} {\enquote {\bibinfo {title} {{Breached Pairing
  Superfluidity: Possible Realization in QCD}},}\ }\href {\doibase
  10.1103/PhysRevLett.91.032001} {\bibfield  {journal} {\bibinfo  {journal}
  {Phys. Rev. Lett.}\ }\textbf {\bibinfo {volume} {91}},\ \bibinfo {pages}
  {032001} (\bibinfo {year} {2003})}\BibitemShut {NoStop}%
\bibitem [{\citenamefont {Moreo}\ \emph
  {et~al.}(2009{\natexlab{a}})\citenamefont {Moreo}, \citenamefont {Daghofer},
  \citenamefont {Riera},\ and\ \citenamefont {Dagotto}}]{amoreo1}%
  \BibitemOpen
  \bibfield  {author} {\bibinfo {author} {\bibfnamefont {A.}~\bibnamefont
  {Moreo}}, \bibinfo {author} {\bibfnamefont {M.}~\bibnamefont {Daghofer}},
  \bibinfo {author} {\bibfnamefont {J.~A.}\ \bibnamefont {Riera}}, \ and\
  \bibinfo {author} {\bibfnamefont {E.}~\bibnamefont {Dagotto}},\ }\bibfield
  {title} {\enquote {\bibinfo {title} {Properties of a two-orbital model for
  oxypnictide superconductors: Magnetic order, ${B}_{2\text{g}}$ spin-singlet
  pairing channel, and its nodal structure},}\ }\href {\doibase
  10.1103/PhysRevB.79.134502} {\bibfield  {journal} {\bibinfo  {journal} {Phys.
  Rev. B}\ }\textbf {\bibinfo {volume} {79}},\ \bibinfo {pages} {134502}
  (\bibinfo {year} {2009}{\natexlab{a}})}\BibitemShut {NoStop}%
\bibitem [{\citenamefont {Moreo}\ \emph
  {et~al.}(2009{\natexlab{b}})\citenamefont {Moreo}, \citenamefont {Daghofer},
  \citenamefont {Nicholson},\ and\ \citenamefont {Dagotto}}]{amoreo2}%
  \BibitemOpen
  \bibfield  {author} {\bibinfo {author} {\bibfnamefont {A.}~\bibnamefont
  {Moreo}}, \bibinfo {author} {\bibfnamefont {M.}~\bibnamefont {Daghofer}},
  \bibinfo {author} {\bibfnamefont {A.}~\bibnamefont {Nicholson}}, \ and\
  \bibinfo {author} {\bibfnamefont {E.}~\bibnamefont {Dagotto}},\ }\bibfield
  {title} {\enquote {\bibinfo {title} {Interband pairing in multiorbital
  systems},}\ }\href {\doibase 10.1103/PhysRevB.80.104507} {\bibfield
  {journal} {\bibinfo  {journal} {Phys. Rev. B}\ }\textbf {\bibinfo {volume}
  {80}},\ \bibinfo {pages} {104507} (\bibinfo {year}
  {2009}{\natexlab{b}})}\BibitemShut {NoStop}%
\bibitem [{\citenamefont {Matt}\ \emph {et~al.}(2018)\citenamefont {Matt},
  \citenamefont {Sutter}, \citenamefont {Cook}, \citenamefont {Sassa},
  \citenamefont {M{\aa}nsson}, \citenamefont {Tjernberg}, \citenamefont {Das},
  \citenamefont {Horio}, \citenamefont {Destraz}, \citenamefont {Fatuzzo},
  \citenamefont {Hauser}, \citenamefont {Shi}, \citenamefont {Kobayashi},
  \citenamefont {Strocov}, \citenamefont {Schmitt}, \citenamefont {Dudin},
  \citenamefont {Hoesch}, \citenamefont {Pyon}, \citenamefont {Takayama},
  \citenamefont {Takagi}, \citenamefont {Lipscombe}, \citenamefont {Hayden},
  \citenamefont {Kurosawa}, \citenamefont {Momono}, \citenamefont {Oda},
  \citenamefont {Neupert},\ and\ \citenamefont {Chang}}]{matt2018}%
  \BibitemOpen
  \bibfield  {author} {\bibinfo {author} {\bibfnamefont {C.~E.}\ \bibnamefont
  {Matt}}, \bibinfo {author} {\bibfnamefont {D.}~\bibnamefont {Sutter}},
  \bibinfo {author} {\bibfnamefont {A.~M.}\ \bibnamefont {Cook}}, \bibinfo
  {author} {\bibfnamefont {Y.}~\bibnamefont {Sassa}}, \bibinfo {author}
  {\bibfnamefont {M.}~\bibnamefont {M{\aa}nsson}}, \bibinfo {author}
  {\bibfnamefont {O.}~\bibnamefont {Tjernberg}}, \bibinfo {author}
  {\bibfnamefont {L.}~\bibnamefont {Das}}, \bibinfo {author} {\bibfnamefont
  {M.}~\bibnamefont {Horio}}, \bibinfo {author} {\bibfnamefont
  {D.}~\bibnamefont {Destraz}}, \bibinfo {author} {\bibfnamefont {C.~G.}\
  \bibnamefont {Fatuzzo}}, \bibinfo {author} {\bibfnamefont {K.}~\bibnamefont
  {Hauser}}, \bibinfo {author} {\bibfnamefont {M.}~\bibnamefont {Shi}},
  \bibinfo {author} {\bibfnamefont {M.}~\bibnamefont {Kobayashi}}, \bibinfo
  {author} {\bibfnamefont {V.~N.}\ \bibnamefont {Strocov}}, \bibinfo {author}
  {\bibfnamefont {T.}~\bibnamefont {Schmitt}}, \bibinfo {author} {\bibfnamefont
  {P.}~\bibnamefont {Dudin}}, \bibinfo {author} {\bibfnamefont
  {M.}~\bibnamefont {Hoesch}}, \bibinfo {author} {\bibfnamefont
  {S.}~\bibnamefont {Pyon}}, \bibinfo {author} {\bibfnamefont {T.}~\bibnamefont
  {Takayama}}, \bibinfo {author} {\bibfnamefont {H.}~\bibnamefont {Takagi}},
  \bibinfo {author} {\bibfnamefont {O.~J.}\ \bibnamefont {Lipscombe}}, \bibinfo
  {author} {\bibfnamefont {S.~M.}\ \bibnamefont {Hayden}}, \bibinfo {author}
  {\bibfnamefont {T.}~\bibnamefont {Kurosawa}}, \bibinfo {author}
  {\bibfnamefont {N.}~\bibnamefont {Momono}}, \bibinfo {author} {\bibfnamefont
  {M.}~\bibnamefont {Oda}}, \bibinfo {author} {\bibfnamefont {T.}~\bibnamefont
  {Neupert}}, \ and\ \bibinfo {author} {\bibfnamefont {J.}~\bibnamefont
  {Chang}},\ }\bibfield  {title} {\enquote {\bibinfo {title} {Direct
  observation of orbital hybridisation in a cuprate superconductor},}\ }\href
  {\doibase 10.1038/s41467-018-03266-0} {\bibfield  {journal} {\bibinfo
  {journal} {Nature Communications}\ }\textbf {\bibinfo {volume} {9}},\
  \bibinfo {pages} {972} (\bibinfo {year} {2018})}\BibitemShut {NoStop}%
\bibitem [{\citenamefont {Tahir-Kheli}(1998)}]{jamil_tahir}%
  \BibitemOpen
  \bibfield  {author} {\bibinfo {author} {\bibfnamefont {J.}~\bibnamefont
  {Tahir-Kheli}},\ }\bibfield  {title} {\enquote {\bibinfo {title} {Interband
  pairing theory of superconductivity},}\ }\href {\doibase
  10.1103/PhysRevB.58.12307} {\bibfield  {journal} {\bibinfo  {journal} {Phys.
  Rev. B}\ }\textbf {\bibinfo {volume} {58}},\ \bibinfo {pages} {12307--12322}
  (\bibinfo {year} {1998})}\BibitemShut {NoStop}%
\bibitem [{\citenamefont {Dolgov}\ \emph {et~al.}(1987)\citenamefont {Dolgov},
  \citenamefont {Fetisov},\ and\ \citenamefont {Khomskii}}]{dolgov}%
  \BibitemOpen
  \bibfield  {author} {\bibinfo {author} {\bibfnamefont {O.V.}\ \bibnamefont
  {Dolgov}}, \bibinfo {author} {\bibfnamefont {E.P.}\ \bibnamefont {Fetisov}},
  \ and\ \bibinfo {author} {\bibfnamefont {D.I.}\ \bibnamefont {Khomskii}},\
  }\bibfield  {title} {\enquote {\bibinfo {title} {Superconductivity of heavy
  fermions in a two-band model},}\ }\href {\doibase
  10.1016/0375-9601(87)90207-6} {\bibfield  {journal} {\bibinfo  {journal}
  {Physics Letters A}\ }\textbf {\bibinfo {volume} {125}},\ \bibinfo {pages}
  {267 -- 270} (\bibinfo {year} {1987})}\BibitemShut {NoStop}%
\bibitem [{\citenamefont {Poncé}\ \emph {et~al.}(2016)\citenamefont {Poncé},
  \citenamefont {Margine}, \citenamefont {Verdi},\ and\ \citenamefont
  {Giustino}}]{ponce}%
  \BibitemOpen
  \bibfield  {author} {\bibinfo {author} {\bibfnamefont {S.}~\bibnamefont
  {Poncé}}, \bibinfo {author} {\bibfnamefont {E.R.}\ \bibnamefont {Margine}},
  \bibinfo {author} {\bibfnamefont {C.}~\bibnamefont {Verdi}}, \ and\ \bibinfo
  {author} {\bibfnamefont {F.}~\bibnamefont {Giustino}},\ }\bibfield  {title}
  {\enquote {\bibinfo {title} {Epw: Electron–phonon coupling, transport and
  superconducting properties using maximally localized wannier functions},}\
  }\href {\doibase 10.1016/j.cpc.2016.07.028} {\bibfield  {journal} {\bibinfo
  {journal} {Computer Physics Communications}\ }\textbf {\bibinfo {volume}
  {209}},\ \bibinfo {pages} {116 -- 133} (\bibinfo {year} {2016})}\BibitemShut
  {NoStop}%
\bibitem [{\citenamefont {Bekaert}\ \emph {et~al.}(2017)\citenamefont
  {Bekaert}, \citenamefont {Bignardi}, \citenamefont {Aperis}, \citenamefont
  {van Abswoude}, \citenamefont {Mattevi}, \citenamefont {Gorovikov},
  \citenamefont {Petaccia}, \citenamefont {Goldoni}, \citenamefont {Partoens},
  \citenamefont {Oppeneer}, \citenamefont {Peeters}, \citenamefont
  {Milo{\v{s}}evi{\'{c}}}, \citenamefont {Rudolf},\ and\ \citenamefont
  {Cepek}}]{jonas}%
  \BibitemOpen
  \bibfield  {author} {\bibinfo {author} {\bibfnamefont {J.}~\bibnamefont
  {Bekaert}}, \bibinfo {author} {\bibfnamefont {L.}~\bibnamefont {Bignardi}},
  \bibinfo {author} {\bibfnamefont {A.}~\bibnamefont {Aperis}}, \bibinfo
  {author} {\bibfnamefont {P.}~\bibnamefont {van Abswoude}}, \bibinfo {author}
  {\bibfnamefont {C.}~\bibnamefont {Mattevi}}, \bibinfo {author} {\bibfnamefont
  {S.}~\bibnamefont {Gorovikov}}, \bibinfo {author} {\bibfnamefont
  {L.}~\bibnamefont {Petaccia}}, \bibinfo {author} {\bibfnamefont
  {A.}~\bibnamefont {Goldoni}}, \bibinfo {author} {\bibfnamefont
  {B.}~\bibnamefont {Partoens}}, \bibinfo {author} {\bibfnamefont {P.~M.}\
  \bibnamefont {Oppeneer}}, \bibinfo {author} {\bibfnamefont {F.~M.}\
  \bibnamefont {Peeters}}, \bibinfo {author} {\bibfnamefont {M.~V.}\
  \bibnamefont {Milo{\v{s}}evi{\'{c}}}}, \bibinfo {author} {\bibfnamefont
  {P.}~\bibnamefont {Rudolf}}, \ and\ \bibinfo {author} {\bibfnamefont
  {C.}~\bibnamefont {Cepek}},\ }\bibfield  {title} {\enquote {\bibinfo {title}
  {{Free surfaces recast superconductivity in few-monolayer MgB$_2$: Combined
  first-principles and ARPES demonstration}},}\ }\href {\doibase
  10.1038/s41598-017-13913-z} {\bibfield  {journal} {\bibinfo  {journal}
  {Scientific Reports}\ }\textbf {\bibinfo {volume} {7}},\ \bibinfo {pages}
  {14458} (\bibinfo {year} {2017})}\BibitemShut {NoStop}%
\bibitem [{\citenamefont {Suhl}\ \emph {et~al.}(1959)\citenamefont {Suhl},
  \citenamefont {Matthias},\ and\ \citenamefont {Walker}}]{suhl_matthias}%
  \BibitemOpen
  \bibfield  {author} {\bibinfo {author} {\bibfnamefont {H.}~\bibnamefont
  {Suhl}}, \bibinfo {author} {\bibfnamefont {B.~T.}\ \bibnamefont {Matthias}},
  \ and\ \bibinfo {author} {\bibfnamefont {L.~R.}\ \bibnamefont {Walker}},\
  }\bibfield  {title} {\enquote {\bibinfo {title} {{Bardeen-Cooper-Schrieffer
  Theory of Superconductivity in the Case of Overlapping Bands}},}\ }\href
  {\doibase 10.1103/PhysRevLett.3.552} {\bibfield  {journal} {\bibinfo
  {journal} {Phys. Rev. Lett.}\ }\textbf {\bibinfo {volume} {3}},\ \bibinfo
  {pages} {552--554} (\bibinfo {year} {1959})}\BibitemShut {NoStop}%
\bibitem [{\citenamefont {Korchorb\'e}\ and\ \citenamefont
  {Palistrant}(1993)}]{korko_palist}%
  \BibitemOpen
  \bibfield  {author} {\bibinfo {author} {\bibfnamefont {F.~G.}\ \bibnamefont
  {Korchorb\'e}}\ and\ \bibinfo {author} {\bibfnamefont {M.~E.}\ \bibnamefont
  {Palistrant}},\ }\bibfield  {title} {\enquote {\bibinfo {title}
  {Superconductivity in a two-band system with low carrier density},}\
  }\href@noop {} {\bibfield  {journal} {\bibinfo  {journal} {Zh. Eksp. Teor.
  Fiz.}\ }\textbf {\bibinfo {volume} {104}},\ \bibinfo {pages} {442--451}
  (\bibinfo {year} {1993})}\BibitemShut {NoStop}%
\bibitem [{\citenamefont {Giubileo}\ \emph {et~al.}(2007)\citenamefont
  {Giubileo}, \citenamefont {Bobba}, \citenamefont {Scarfato}, \citenamefont
  {Cucolo}, \citenamefont {Kohen}, \citenamefont {Roditchev}, \citenamefont
  {Zhigadlo},\ and\ \citenamefont {Karpinski}}]{roditchev_prox_effect}%
  \BibitemOpen
  \bibfield  {author} {\bibinfo {author} {\bibfnamefont {F.}~\bibnamefont
  {Giubileo}}, \bibinfo {author} {\bibfnamefont {F.}~\bibnamefont {Bobba}},
  \bibinfo {author} {\bibfnamefont {A.}~\bibnamefont {Scarfato}}, \bibinfo
  {author} {\bibfnamefont {A.~M.}\ \bibnamefont {Cucolo}}, \bibinfo {author}
  {\bibfnamefont {A.}~\bibnamefont {Kohen}}, \bibinfo {author} {\bibfnamefont
  {D.}~\bibnamefont {Roditchev}}, \bibinfo {author} {\bibfnamefont {N.~D.}\
  \bibnamefont {Zhigadlo}}, \ and\ \bibinfo {author} {\bibfnamefont
  {J.}~\bibnamefont {Karpinski}},\ }\bibfield  {title} {\enquote {\bibinfo
  {title} {{Local tunneling study of three-dimensional order parameter in the
  $\pi$ band of Al-doped MgB$_2$ single crystals}},}\ }\href {\doibase
  10.1103/PhysRevB.76.024507} {\bibfield  {journal} {\bibinfo  {journal} {Phys.
  Rev. B}\ }\textbf {\bibinfo {volume} {76}},\ \bibinfo {pages} {024507}
  (\bibinfo {year} {2007})}\BibitemShut {NoStop}%
\bibitem [{\citenamefont {Giubileo}\ \emph {et~al.}(2001)\citenamefont
  {Giubileo}, \citenamefont {Roditchev}, \citenamefont {Sacks}, \citenamefont
  {Lamy}, \citenamefont {Thanh}, \citenamefont {Klein}, \citenamefont
  {Miraglia}, \citenamefont {Fruchart}, \citenamefont {Marcus},\ and\
  \citenamefont {Monod}}]{roditchev_prox_effect2}%
  \BibitemOpen
  \bibfield  {author} {\bibinfo {author} {\bibfnamefont {F.}~\bibnamefont
  {Giubileo}}, \bibinfo {author} {\bibfnamefont {D.}~\bibnamefont {Roditchev}},
  \bibinfo {author} {\bibfnamefont {W.}~\bibnamefont {Sacks}}, \bibinfo
  {author} {\bibfnamefont {R.}~\bibnamefont {Lamy}}, \bibinfo {author}
  {\bibfnamefont {D.~X.}\ \bibnamefont {Thanh}}, \bibinfo {author}
  {\bibfnamefont {J.}~\bibnamefont {Klein}}, \bibinfo {author} {\bibfnamefont
  {S.}~\bibnamefont {Miraglia}}, \bibinfo {author} {\bibfnamefont
  {D.}~\bibnamefont {Fruchart}}, \bibinfo {author} {\bibfnamefont
  {J.}~\bibnamefont {Marcus}}, \ and\ \bibinfo {author} {\bibfnamefont {Ph.}\
  \bibnamefont {Monod}},\ }\bibfield  {title} {\enquote {\bibinfo {title}
  {{Two-Gap State Density in MgB$_2$: A True Bulk Property Or A Proximity
  Effect?}}}\ }\href {\doibase 10.1103/PhysRevLett.87.177008} {\bibfield
  {journal} {\bibinfo  {journal} {Phys. Rev. Lett.}\ }\textbf {\bibinfo
  {volume} {87}},\ \bibinfo {pages} {177008} (\bibinfo {year}
  {2001})}\BibitemShut {NoStop}%
\bibitem [{\citenamefont {Damascelli}\ \emph {et~al.}(2003)\citenamefont
  {Damascelli}, \citenamefont {Hussain},\ and\ \citenamefont {Shen}}]{dama}%
  \BibitemOpen
  \bibfield  {author} {\bibinfo {author} {\bibfnamefont {Andrea}\ \bibnamefont
  {Damascelli}}, \bibinfo {author} {\bibfnamefont {Zahid}\ \bibnamefont
  {Hussain}}, \ and\ \bibinfo {author} {\bibfnamefont {Zhi-Xun}\ \bibnamefont
  {Shen}},\ }\bibfield  {title} {\enquote {\bibinfo {title} {Angle-resolved
  photoemission studies of the cuprate superconductors},}\ }\href {\doibase
  10.1103/RevModPhys.75.473} {\bibfield  {journal} {\bibinfo  {journal} {Rev.
  Mod. Phys.}\ }\textbf {\bibinfo {volume} {75}},\ \bibinfo {pages} {473--541}
  (\bibinfo {year} {2003})}\BibitemShut {NoStop}%
\bibitem [{\citenamefont {Souma}\ \emph {et~al.}(2003)\citenamefont {Souma},
  \citenamefont {Machida}, \citenamefont {Sato}, \citenamefont {Takahashi},
  \citenamefont {Matsui}, \citenamefont {Wang}, \citenamefont {Ding},
  \citenamefont {Kaminski}, \citenamefont {Campuzano}, \citenamefont {Sasaki},\
  and\ \citenamefont {Kadowaki}}]{soumamgb2}%
  \BibitemOpen
  \bibfield  {author} {\bibinfo {author} {\bibfnamefont {S.}~\bibnamefont
  {Souma}}, \bibinfo {author} {\bibfnamefont {Y.}~\bibnamefont {Machida}},
  \bibinfo {author} {\bibfnamefont {T.}~\bibnamefont {Sato}}, \bibinfo {author}
  {\bibfnamefont {T.}~\bibnamefont {Takahashi}}, \bibinfo {author}
  {\bibfnamefont {H.}~\bibnamefont {Matsui}}, \bibinfo {author} {\bibfnamefont
  {S.-C.}\ \bibnamefont {Wang}}, \bibinfo {author} {\bibfnamefont
  {H.}~\bibnamefont {Ding}}, \bibinfo {author} {\bibfnamefont {A.}~\bibnamefont
  {Kaminski}}, \bibinfo {author} {\bibfnamefont {J.~C.}\ \bibnamefont
  {Campuzano}}, \bibinfo {author} {\bibfnamefont {S.}~\bibnamefont {Sasaki}}, \
  and\ \bibinfo {author} {\bibfnamefont {K.}~\bibnamefont {Kadowaki}},\
  }\bibfield  {title} {\enquote {\bibinfo {title} {{The origin of multiple
  superconducting gaps in MgB$_2$}},}\ }\href
  {https://doi.org/10.1038/nature01619 http://10.0.4.14/nature01619} {\bibfield
   {journal} {\bibinfo  {journal} {Nature}\ }\textbf {\bibinfo {volume}
  {423}},\ \bibinfo {pages} {65} (\bibinfo {year} {2003})}\BibitemShut
  {NoStop}%
\bibitem [{\citenamefont {Kuzmichev}\ \emph {et~al.}(2014)\citenamefont
  {Kuzmichev}, \citenamefont {Kuzmicheva},\ and\ \citenamefont
  {Tchesnokov}}]{kuzmichev}%
  \BibitemOpen
  \bibfield  {author} {\bibinfo {author} {\bibfnamefont {S.~A.}\ \bibnamefont
  {Kuzmichev}}, \bibinfo {author} {\bibfnamefont {T.~E.}\ \bibnamefont
  {Kuzmicheva}}, \ and\ \bibinfo {author} {\bibfnamefont {S.~N.}\ \bibnamefont
  {Tchesnokov}},\ }\bibfield  {title} {\enquote {\bibinfo {title}
  {{Determination of the electron-phonon coupling constants from the
  experimental temperature dependences of superconducting gaps in MgB$_2$}},}\
  }\href {\doibase 10.1134/S0021364014050129} {\bibfield  {journal} {\bibinfo
  {journal} {JETP Letters}\ }\textbf {\bibinfo {volume} {99}},\ \bibinfo
  {pages} {295--302} (\bibinfo {year} {2014})}\BibitemShut {NoStop}%
\bibitem [{\citenamefont {Choi}\ \emph {et~al.}(2002)\citenamefont {Choi},
  \citenamefont {Roundy}, \citenamefont {Sun}, \citenamefont {Cohen},\ and\
  \citenamefont {Louie}}]{choi_mgb2}%
  \BibitemOpen
  \bibfield  {author} {\bibinfo {author} {\bibfnamefont {H.~J.}\ \bibnamefont
  {Choi}}, \bibinfo {author} {\bibfnamefont {D.}~\bibnamefont {Roundy}},
  \bibinfo {author} {\bibfnamefont {H.}~\bibnamefont {Sun}}, \bibinfo {author}
  {\bibfnamefont {M.~L.}\ \bibnamefont {Cohen}}, \ and\ \bibinfo {author}
  {\bibfnamefont {S.~G.}\ \bibnamefont {Louie}},\ }\bibfield  {title} {\enquote
  {\bibinfo {title} {{First-principles calculation of the superconducting
  transition in MgB$_2$ within the anisotropic Eliashberg formalism}},}\ }\href
  {\doibase 10.1103/PhysRevB.66.020513} {\bibfield  {journal} {\bibinfo
  {journal} {Phys. Rev. B}\ }\textbf {\bibinfo {volume} {66}},\ \bibinfo
  {pages} {020513} (\bibinfo {year} {2002})}\BibitemShut {NoStop}%
\bibitem [{\citenamefont {Aperis}\ \emph {et~al.}(2015)\citenamefont {Aperis},
  \citenamefont {Maldonado},\ and\ \citenamefont {Oppeneer}}]{aperis_mgb2}%
  \BibitemOpen
  \bibfield  {author} {\bibinfo {author} {\bibfnamefont {A.}~\bibnamefont
  {Aperis}}, \bibinfo {author} {\bibfnamefont {P.}~\bibnamefont {Maldonado}}, \
  and\ \bibinfo {author} {\bibfnamefont {P.~M.}\ \bibnamefont {Oppeneer}},\
  }\bibfield  {title} {\enquote {\bibinfo {title} {{Ab initio theory of
  magnetic-field-induced odd-frequency two-band superconductivity in
  MgB$_2$}},}\ }\href {\doibase 10.1103/PhysRevB.92.054516} {\bibfield
  {journal} {\bibinfo  {journal} {Phys. Rev. B}\ }\textbf {\bibinfo {volume}
  {92}},\ \bibinfo {pages} {054516} (\bibinfo {year} {2015})}\BibitemShut
  {NoStop}%
\bibitem [{\citenamefont {Mazin}\ \emph {et~al.}(2008)\citenamefont {Mazin},
  \citenamefont {Singh}, \citenamefont {Johannes},\ and\ \citenamefont
  {Du}}]{mazin}%
  \BibitemOpen
  \bibfield  {author} {\bibinfo {author} {\bibfnamefont {I.~I.}\ \bibnamefont
  {Mazin}}, \bibinfo {author} {\bibfnamefont {D.~J.}\ \bibnamefont {Singh}},
  \bibinfo {author} {\bibfnamefont {M.~D.}\ \bibnamefont {Johannes}}, \ and\
  \bibinfo {author} {\bibfnamefont {M.~H.}\ \bibnamefont {Du}},\ }\bibfield
  {title} {\enquote {\bibinfo {title} {{Unconventional Superconductivity with a
  Sign Reversal in the Order Parameter of
  ${\mathrm{LaFeAsO}}_{1\ensuremath{-}x}{\mathrm{F}}_{x}$}},}\ }\href {\doibase
  10.1103/PhysRevLett.101.057003} {\bibfield  {journal} {\bibinfo  {journal}
  {Phys. Rev. Lett.}\ }\textbf {\bibinfo {volume} {101}},\ \bibinfo {pages}
  {057003} (\bibinfo {year} {2008})}\BibitemShut {NoStop}%
\bibitem [{\citenamefont {Lu}\ \emph {et~al.}(2012)\citenamefont {Lu},
  \citenamefont {Fang}, \citenamefont {Tsai}, \citenamefont {Jiang},\ and\
  \citenamefont {Hu}}]{xiaoli}%
  \BibitemOpen
  \bibfield  {author} {\bibinfo {author} {\bibfnamefont {X.}~\bibnamefont
  {Lu}}, \bibinfo {author} {\bibfnamefont {C.}~\bibnamefont {Fang}}, \bibinfo
  {author} {\bibfnamefont {W-F}\ \bibnamefont {Tsai}}, \bibinfo {author}
  {\bibfnamefont {Y.}~\bibnamefont {Jiang}}, \ and\ \bibinfo {author}
  {\bibfnamefont {J.}~\bibnamefont {Hu}},\ }\bibfield  {title} {\enquote
  {\bibinfo {title} {{$s$-wave superconductivity with orbital-dependent sign
  change in checkerboard models of iron-based superconductors}},}\ }\href
  {\doibase 10.1103/PhysRevB.85.054505} {\bibfield  {journal} {\bibinfo
  {journal} {Phys. Rev. B}\ }\textbf {\bibinfo {volume} {85}},\ \bibinfo
  {pages} {054505} (\bibinfo {year} {2012})}\BibitemShut {NoStop}%
\bibitem [{\citenamefont {Zhang}\ \emph {et~al.}(2014)\citenamefont {Zhang},
  \citenamefont {Richard}, \citenamefont {Qian}, \citenamefont {Shi},
  \citenamefont {Ma}, \citenamefont {Zeng}, \citenamefont {Wang}, \citenamefont
  {Rienks}, \citenamefont {Zhang}, \citenamefont {Dai}, \citenamefont {You},
  \citenamefont {Weng}, \citenamefont {Wu}, \citenamefont {Hu},\ and\
  \citenamefont {Ding}}]{pzhang}%
  \BibitemOpen
  \bibfield  {author} {\bibinfo {author} {\bibfnamefont {P.}~\bibnamefont
  {Zhang}}, \bibinfo {author} {\bibfnamefont {P.}~\bibnamefont {Richard}},
  \bibinfo {author} {\bibfnamefont {T.}~\bibnamefont {Qian}}, \bibinfo {author}
  {\bibfnamefont {X.}~\bibnamefont {Shi}}, \bibinfo {author} {\bibfnamefont
  {J.}~\bibnamefont {Ma}}, \bibinfo {author} {\bibfnamefont {L.-K.}\
  \bibnamefont {Zeng}}, \bibinfo {author} {\bibfnamefont {X.-P.}\ \bibnamefont
  {Wang}}, \bibinfo {author} {\bibfnamefont {E.}~\bibnamefont {Rienks}},
  \bibinfo {author} {\bibfnamefont {C.-L.}\ \bibnamefont {Zhang}}, \bibinfo
  {author} {\bibfnamefont {P.}~\bibnamefont {Dai}}, \bibinfo {author}
  {\bibfnamefont {Y.-Z.}\ \bibnamefont {You}}, \bibinfo {author} {\bibfnamefont
  {Z.-Y.}\ \bibnamefont {Weng}}, \bibinfo {author} {\bibfnamefont {X.-X.}\
  \bibnamefont {Wu}}, \bibinfo {author} {\bibfnamefont {J.~P.}\ \bibnamefont
  {Hu}}, \ and\ \bibinfo {author} {\bibfnamefont {H.}~\bibnamefont {Ding}},\
  }\bibfield  {title} {\enquote {\bibinfo {title} {{Observation of
  Momentum-Confined In-Gap Impurity State in Ba$_{0.6}$K$_{0.4}$Fe$_2$As$_2$:
  Evidence for Antiphase s$_\pm$ Pairing}},}\ }\href {\doibase
  10.1103/PhysRevX.4.031001} {\bibfield  {journal} {\bibinfo  {journal} {Phys.
  Rev. X}\ }\textbf {\bibinfo {volume} {4}},\ \bibinfo {pages} {031001}
  (\bibinfo {year} {2014})}\BibitemShut {NoStop}%
\bibitem [{\citenamefont {Yin}\ \emph {et~al.}(2014)\citenamefont {Yin},
  \citenamefont {Haule},\ and\ \citenamefont {Kotliar}}]{yin2014}%
  \BibitemOpen
  \bibfield  {author} {\bibinfo {author} {\bibfnamefont {Z.~P.}\ \bibnamefont
  {Yin}}, \bibinfo {author} {\bibfnamefont {K.}~\bibnamefont {Haule}}, \ and\
  \bibinfo {author} {\bibfnamefont {G.}~\bibnamefont {Kotliar}},\ }\bibfield
  {title} {\enquote {\bibinfo {title} {{Spin dynamics and orbital-antiphase
  pairing symmetry in iron-based superconductors}},}\ }\href
  {https://doi.org/10.1038/nphys3116 http://10.0.4.14/nphys3116
  https://www.nature.com/articles/nphys3116{\#}supplementary-information}
  {\bibfield  {journal} {\bibinfo  {journal} {Nature Physics}\ }\textbf
  {\bibinfo {volume} {10}},\ \bibinfo {pages} {845} (\bibinfo {year}
  {2014})}\BibitemShut {NoStop}%
\bibitem [{\citenamefont {Ding}\ \emph {et~al.}(2011)\citenamefont {Ding},
  \citenamefont {Nakayama}, \citenamefont {Richard}, \citenamefont {Souma},
  \citenamefont {Sato}, \citenamefont {Takahashi}, \citenamefont {Neupane},
  \citenamefont {Xu}, \citenamefont {Pan}, \citenamefont {Fedorov},
  \citenamefont {Wang}, \citenamefont {Dai}, \citenamefont {Fang},
  \citenamefont {Chen}, \citenamefont {Luo},\ and\ \citenamefont
  {Wang}}]{ding2}%
  \BibitemOpen
  \bibfield  {author} {\bibinfo {author} {\bibfnamefont {H.}~\bibnamefont
  {Ding}}, \bibinfo {author} {\bibfnamefont {K.}~\bibnamefont {Nakayama}},
  \bibinfo {author} {\bibfnamefont {P.}~\bibnamefont {Richard}}, \bibinfo
  {author} {\bibfnamefont {S.}~\bibnamefont {Souma}}, \bibinfo {author}
  {\bibfnamefont {T.}~\bibnamefont {Sato}}, \bibinfo {author} {\bibfnamefont
  {T.}~\bibnamefont {Takahashi}}, \bibinfo {author} {\bibfnamefont
  {M.}~\bibnamefont {Neupane}}, \bibinfo {author} {\bibfnamefont {Y-M}\
  \bibnamefont {Xu}}, \bibinfo {author} {\bibfnamefont {Z-H}\ \bibnamefont
  {Pan}}, \bibinfo {author} {\bibfnamefont {A.~V.}\ \bibnamefont {Fedorov}},
  \bibinfo {author} {\bibfnamefont {Z.}~\bibnamefont {Wang}}, \bibinfo {author}
  {\bibfnamefont {X.}~\bibnamefont {Dai}}, \bibinfo {author} {\bibfnamefont
  {Z.}~\bibnamefont {Fang}}, \bibinfo {author} {\bibfnamefont {G.~F.}\
  \bibnamefont {Chen}}, \bibinfo {author} {\bibfnamefont {J.~L.}\ \bibnamefont
  {Luo}}, \ and\ \bibinfo {author} {\bibfnamefont {N.~L.}\ \bibnamefont
  {Wang}},\ }\bibfield  {title} {\enquote {\bibinfo {title} {{Electronic
  structure of optimally doped pnictide Ba$_{0.6}$K$_{0.4}$Fe$_2$As$_2$: a
  comprehensive angle-resolved photoemission spectroscopy investigation}},}\
  }\href {\doibase 10.1088/0953-8984/23/13/135701} {\bibfield  {journal}
  {\bibinfo  {journal} {Journal of Physics: Condensed Matter}\ }\textbf
  {\bibinfo {volume} {23}},\ \bibinfo {pages} {135701} (\bibinfo {year}
  {2011})}\BibitemShut {NoStop}%
\bibitem [{\citenamefont {Salovich}\ \emph {et~al.}(2013)\citenamefont
  {Salovich}, \citenamefont {Kim}, \citenamefont {Ghosh}, \citenamefont
  {Giannetta}, \citenamefont {Kwok}, \citenamefont {Welp}, \citenamefont
  {Shen}, \citenamefont {Zhu}, \citenamefont {Wen}, \citenamefont {Tanatar},\
  and\ \citenamefont {Prozorov}}]{salovich}%
  \BibitemOpen
  \bibfield  {author} {\bibinfo {author} {\bibfnamefont {N.~W.}\ \bibnamefont
  {Salovich}}, \bibinfo {author} {\bibfnamefont {H.}~\bibnamefont {Kim}},
  \bibinfo {author} {\bibfnamefont {A.~K.}\ \bibnamefont {Ghosh}}, \bibinfo
  {author} {\bibfnamefont {R.~W.}\ \bibnamefont {Giannetta}}, \bibinfo {author}
  {\bibfnamefont {W.}~\bibnamefont {Kwok}}, \bibinfo {author} {\bibfnamefont
  {U.}~\bibnamefont {Welp}}, \bibinfo {author} {\bibfnamefont {B.}~\bibnamefont
  {Shen}}, \bibinfo {author} {\bibfnamefont {S.}~\bibnamefont {Zhu}}, \bibinfo
  {author} {\bibfnamefont {H.-H.}\ \bibnamefont {Wen}}, \bibinfo {author}
  {\bibfnamefont {M.~A.}\ \bibnamefont {Tanatar}}, \ and\ \bibinfo {author}
  {\bibfnamefont {R.}~\bibnamefont {Prozorov}},\ }\bibfield  {title} {\enquote
  {\bibinfo {title} {{Effect of heavy-ion irradiation on superconductivity in
  Ba${}_{0.6}$K${}_{0.4}$Fe${}_{2}$As${}_{2}$}},}\ }\href {\doibase
  10.1103/PhysRevB.87.180502} {\bibfield  {journal} {\bibinfo  {journal} {Phys.
  Rev. B}\ }\textbf {\bibinfo {volume} {87}},\ \bibinfo {pages} {180502}
  (\bibinfo {year} {2013})}\BibitemShut {NoStop}%
\bibitem [{\citenamefont {Stanev}\ and\ \citenamefont
  {Koshelev}(2012)}]{stanev}%
  \BibitemOpen
  \bibfield  {author} {\bibinfo {author} {\bibfnamefont {V.~G.}\ \bibnamefont
  {Stanev}}\ and\ \bibinfo {author} {\bibfnamefont {A.~E.}\ \bibnamefont
  {Koshelev}},\ }\bibfield  {title} {\enquote {\bibinfo {title} {{Anomalous
  proximity effects at the interface of $s$ and ${s}_{\pm}$
  superconductors}},}\ }\href {\doibase 10.1103/PhysRevB.86.174515} {\bibfield
  {journal} {\bibinfo  {journal} {Phys. Rev. B}\ }\textbf {\bibinfo {volume}
  {86}},\ \bibinfo {pages} {174515} (\bibinfo {year} {2012})}\BibitemShut
  {NoStop}%
\bibitem [{\citenamefont {Orlova}\ \emph {et~al.}(2013)\citenamefont {Orlova},
  \citenamefont {Shanenko}, \citenamefont {Milo\v{s}evi\'c}, \citenamefont
  {Peeters}, \citenamefont {Vagov},\ and\ \citenamefont {Axt}}]{n_orlova}%
  \BibitemOpen
  \bibfield  {author} {\bibinfo {author} {\bibfnamefont {N.~V.}\ \bibnamefont
  {Orlova}}, \bibinfo {author} {\bibfnamefont {A.~A.}\ \bibnamefont
  {Shanenko}}, \bibinfo {author} {\bibfnamefont {M.~V.}\ \bibnamefont
  {Milo\v{s}evi\'c}}, \bibinfo {author} {\bibfnamefont {F.~M.}\ \bibnamefont
  {Peeters}}, \bibinfo {author} {\bibfnamefont {A.~V.}\ \bibnamefont {Vagov}},
  \ and\ \bibinfo {author} {\bibfnamefont {V.~M.}\ \bibnamefont {Axt}},\
  }\bibfield  {title} {\enquote {\bibinfo {title} {Ginzburg-landau theory for
  multiband superconductors: Microscopic derivation},}\ }\href {\doibase
  10.1103/PhysRevB.87.134510} {\bibfield  {journal} {\bibinfo  {journal} {Phys.
  Rev. B}\ }\textbf {\bibinfo {volume} {87}},\ \bibinfo {pages} {134510}
  (\bibinfo {year} {2013})}\BibitemShut {NoStop}%
\bibitem [{\citenamefont {Brendan}\ and\ \citenamefont
  {Mukunda}(2013)}]{trsb_3band}%
  \BibitemOpen
  \bibfield  {author} {\bibinfo {author} {\bibfnamefont {J.~W.}\ \bibnamefont
  {Brendan}}\ and\ \bibinfo {author} {\bibfnamefont {P.~D.}\ \bibnamefont
  {Mukunda}},\ }\bibfield  {title} {\enquote {\bibinfo {title}
  {Time-reversal-symmetry-broken state in the {BCS} formalism for a multi-band
  superconductor},}\ }\href {\doibase 10.1088/0953-8984/25/42/425702}
  {\bibfield  {journal} {\bibinfo  {journal} {J. Phys. Condens. Matter}\
  }\textbf {\bibinfo {volume} {25}},\ \bibinfo {pages} {425702} (\bibinfo
  {year} {2013})}\BibitemShut {NoStop}%
\bibitem [{\citenamefont {Garaud}\ \emph {et~al.}(2011)\citenamefont {Garaud},
  \citenamefont {Carlstr\"om},\ and\ \citenamefont {Babaev}}]{3b_babaev}%
  \BibitemOpen
  \bibfield  {author} {\bibinfo {author} {\bibfnamefont {J.}~\bibnamefont
  {Garaud}}, \bibinfo {author} {\bibfnamefont {J.}~\bibnamefont {Carlstr\"om}},
  \ and\ \bibinfo {author} {\bibfnamefont {E.}~\bibnamefont {Babaev}},\
  }\bibfield  {title} {\enquote {\bibinfo {title} {Topological solitons in
  three-band superconductors with broken time reversal symmetry},}\ }\href
  {\doibase 10.1103/PhysRevLett.107.197001} {\bibfield  {journal} {\bibinfo
  {journal} {Phys. Rev. Lett.}\ }\textbf {\bibinfo {volume} {107}},\ \bibinfo
  {pages} {197001} (\bibinfo {year} {2011})}\BibitemShut {NoStop}%
\bibitem [{\citenamefont {Garaud}\ \emph {et~al.}(2013)\citenamefont {Garaud},
  \citenamefont {Carlstr\"om}, \citenamefont {Babaev},\ and\ \citenamefont
  {Speight}}]{skyrm_garaud}%
  \BibitemOpen
  \bibfield  {author} {\bibinfo {author} {\bibfnamefont {J.}~\bibnamefont
  {Garaud}}, \bibinfo {author} {\bibfnamefont {J.}~\bibnamefont {Carlstr\"om}},
  \bibinfo {author} {\bibfnamefont {E.}~\bibnamefont {Babaev}}, \ and\ \bibinfo
  {author} {\bibfnamefont {M.}~\bibnamefont {Speight}},\ }\bibfield  {title}
  {\enquote {\bibinfo {title} {{Chiral $\mathbb{C}{P}^{2}$ skyrmions in
  three-band superconductors}},}\ }\href {\doibase 10.1103/PhysRevB.87.014507}
  {\bibfield  {journal} {\bibinfo  {journal} {Phys. Rev. B}\ }\textbf {\bibinfo
  {volume} {87}},\ \bibinfo {pages} {014507} (\bibinfo {year}
  {2013})}\BibitemShut {NoStop}%
\bibitem [{\citenamefont {Orlova}\ \emph {et~al.}(2016)\citenamefont {Orlova},
  \citenamefont {Kuopanportti},\ and\ \citenamefont {Milo\ifmmode
  \check{s}\else \v{s}\fi{}evi\ifmmode~\acute{c}\else
  \'{c}\fi{}}}]{skyr_orlova}%
  \BibitemOpen
  \bibfield  {author} {\bibinfo {author} {\bibfnamefont {N.~V.}\ \bibnamefont
  {Orlova}}, \bibinfo {author} {\bibfnamefont {P.}~\bibnamefont
  {Kuopanportti}}, \ and\ \bibinfo {author} {\bibfnamefont {M.~V.}\
  \bibnamefont {Milo\ifmmode \check{s}\else
  \v{s}\fi{}evi\ifmmode~\acute{c}\else \'{c}\fi{}}},\ }\bibfield  {title}
  {\enquote {\bibinfo {title} {Skyrmionic vortex lattices in coherently coupled
  three-component bose-einstein condensates},}\ }\href {\doibase
  10.1103/PhysRevA.94.023617} {\bibfield  {journal} {\bibinfo  {journal} {Phys.
  Rev. A}\ }\textbf {\bibinfo {volume} {94}},\ \bibinfo {pages} {023617}
  (\bibinfo {year} {2016})}\BibitemShut {NoStop}%
\bibitem [{\citenamefont {Garaud}\ \emph {et~al.}(2017)\citenamefont {Garaud},
  \citenamefont {Silaev},\ and\ \citenamefont {Babaev}}]{change_vortex_core}%
  \BibitemOpen
  \bibfield  {author} {\bibinfo {author} {\bibfnamefont {J.}~\bibnamefont
  {Garaud}}, \bibinfo {author} {\bibfnamefont {M.}~\bibnamefont {Silaev}}, \
  and\ \bibinfo {author} {\bibfnamefont {E.}~\bibnamefont {Babaev}},\
  }\bibfield  {title} {\enquote {\bibinfo {title} {{Change of the vortex core
  structure in two-band superconductors at the impurity-scattering-driven
  ${s}_{\pm}/{s}_{++}$ crossover}},}\ }\href {\doibase
  10.1103/PhysRevB.96.140503} {\bibfield  {journal} {\bibinfo  {journal} {Phys.
  Rev. B}\ }\textbf {\bibinfo {volume} {96}},\ \bibinfo {pages} {140503}
  (\bibinfo {year} {2017})}\BibitemShut {NoStop}%
\bibitem [{\citenamefont {Wang}\ \emph {et~al.}(2018)\citenamefont {Wang},
  \citenamefont {Kong}, \citenamefont {Fan}, \citenamefont {Chen},
  \citenamefont {Zhu}, \citenamefont {Liu}, \citenamefont {Cao}, \citenamefont
  {Sun}, \citenamefont {Du}, \citenamefont {Schneeloch}, \citenamefont {Zhong},
  \citenamefont {Gu}, \citenamefont {Fu}, \citenamefont {Ding},\ and\
  \citenamefont {Gao}}]{wang}%
  \BibitemOpen
  \bibfield  {author} {\bibinfo {author} {\bibfnamefont {D.}~\bibnamefont
  {Wang}}, \bibinfo {author} {\bibfnamefont {L.}~\bibnamefont {Kong}}, \bibinfo
  {author} {\bibfnamefont {P.}~\bibnamefont {Fan}}, \bibinfo {author}
  {\bibfnamefont {H.}~\bibnamefont {Chen}}, \bibinfo {author} {\bibfnamefont
  {S.}~\bibnamefont {Zhu}}, \bibinfo {author} {\bibfnamefont {W.}~\bibnamefont
  {Liu}}, \bibinfo {author} {\bibfnamefont {L.}~\bibnamefont {Cao}}, \bibinfo
  {author} {\bibfnamefont {Y.}~\bibnamefont {Sun}}, \bibinfo {author}
  {\bibfnamefont {S.}~\bibnamefont {Du}}, \bibinfo {author} {\bibfnamefont
  {J.}~\bibnamefont {Schneeloch}}, \bibinfo {author} {\bibfnamefont
  {R.}~\bibnamefont {Zhong}}, \bibinfo {author} {\bibfnamefont
  {G.}~\bibnamefont {Gu}}, \bibinfo {author} {\bibfnamefont {L.}~\bibnamefont
  {Fu}}, \bibinfo {author} {\bibfnamefont {H.}~\bibnamefont {Ding}}, \ and\
  \bibinfo {author} {\bibfnamefont {H-J}\ \bibnamefont {Gao}},\ }\bibfield
  {title} {\enquote {\bibinfo {title} {Evidence for majorana bound states in an
  iron-based superconductor},}\ }\href {\doibase 10.1126/science.aao1797}
  {\bibfield  {journal} {\bibinfo  {journal} {Science}\ }\textbf {\bibinfo
  {volume} {362}},\ \bibinfo {pages} {333--335} (\bibinfo {year}
  {2018})}\BibitemShut {NoStop}%
\bibitem [{\citenamefont {Yin}\ \emph {et~al.}(2015)\citenamefont {Yin},
  \citenamefont {Wu}, \citenamefont {Wang}, \citenamefont {Ye}, \citenamefont
  {Gong}, \citenamefont {Hou}, \citenamefont {Shan}, \citenamefont {Li},
  \citenamefont {Liang}, \citenamefont {Wu}, \citenamefont {Li}, \citenamefont
  {Ting}, \citenamefont {Wang}, \citenamefont {Hu}, \citenamefont {Hor},
  \citenamefont {Ding},\ and\ \citenamefont {Pan}}]{yin2015}%
  \BibitemOpen
  \bibfield  {author} {\bibinfo {author} {\bibfnamefont {J-X.}\ \bibnamefont
  {Yin}}, \bibinfo {author} {\bibfnamefont {Z.}~\bibnamefont {Wu}}, \bibinfo
  {author} {\bibfnamefont {J-H.}\ \bibnamefont {Wang}}, \bibinfo {author}
  {\bibfnamefont {Z-Y.}\ \bibnamefont {Ye}}, \bibinfo {author} {\bibfnamefont
  {Jing}\ \bibnamefont {Gong}}, \bibinfo {author} {\bibfnamefont {X-Y.}\
  \bibnamefont {Hou}}, \bibinfo {author} {\bibfnamefont {Lei}\ \bibnamefont
  {Shan}}, \bibinfo {author} {\bibfnamefont {Ang}\ \bibnamefont {Li}}, \bibinfo
  {author} {\bibfnamefont {X-J.}\ \bibnamefont {Liang}}, \bibinfo {author}
  {\bibfnamefont {X-X.}\ \bibnamefont {Wu}}, \bibinfo {author} {\bibfnamefont
  {Jian}\ \bibnamefont {Li}}, \bibinfo {author} {\bibfnamefont {C-S.}\
  \bibnamefont {Ting}}, \bibinfo {author} {\bibfnamefont {Z-Q.}\ \bibnamefont
  {Wang}}, \bibinfo {author} {\bibfnamefont {J-P.}\ \bibnamefont {Hu}},
  \bibinfo {author} {\bibfnamefont {P-H.}\ \bibnamefont {Hor}}, \bibinfo
  {author} {\bibfnamefont {H.}~\bibnamefont {Ding}}, \ and\ \bibinfo {author}
  {\bibfnamefont {S.~H.}\ \bibnamefont {Pan}},\ }\bibfield  {title} {\enquote
  {\bibinfo {title} {{Observation of a robust zero-energy bound state in
  iron-based superconductor Fe(Te,Se)}},}\ }\href
  {https://doi.org/10.1038/nphys3371 http://10.0.4.14/nphys3371
  https://www.nature.com/articles/nphys3371{\#}supplementary-information}
  {\bibfield  {journal} {\bibinfo  {journal} {Nature Physics}\ }\textbf
  {\bibinfo {volume} {11}},\ \bibinfo {pages} {543} (\bibinfo {year}
  {2015})}\BibitemShut {NoStop}%
\bibitem [{\citenamefont {Gonnelli}\ \emph {et~al.}(2006)\citenamefont
  {Gonnelli}, \citenamefont {Daghero}, \citenamefont {Ummarino}, \citenamefont
  {Calzolari}, \citenamefont {Tortello}, \citenamefont {Stepanov},
  \citenamefont {Zhigadlo}, \citenamefont {Rogacki}, \citenamefont {Karpinski},
  \citenamefont {Bernardini},\ and\ \citenamefont {Massidda}}]{gonelli}%
  \BibitemOpen
  \bibfield  {author} {\bibinfo {author} {\bibfnamefont {R.~S.}\ \bibnamefont
  {Gonnelli}}, \bibinfo {author} {\bibfnamefont {D.}~\bibnamefont {Daghero}},
  \bibinfo {author} {\bibfnamefont {G.~A.}\ \bibnamefont {Ummarino}}, \bibinfo
  {author} {\bibfnamefont {A.}~\bibnamefont {Calzolari}}, \bibinfo {author}
  {\bibfnamefont {M.}~\bibnamefont {Tortello}}, \bibinfo {author}
  {\bibfnamefont {V.~A.}\ \bibnamefont {Stepanov}}, \bibinfo {author}
  {\bibfnamefont {N.~D.}\ \bibnamefont {Zhigadlo}}, \bibinfo {author}
  {\bibfnamefont {K.}~\bibnamefont {Rogacki}}, \bibinfo {author} {\bibfnamefont
  {J.}~\bibnamefont {Karpinski}}, \bibinfo {author} {\bibfnamefont
  {F.}~\bibnamefont {Bernardini}}, \ and\ \bibinfo {author} {\bibfnamefont
  {S.}~\bibnamefont {Massidda}},\ }\bibfield  {title} {\enquote {\bibinfo
  {title} {{Effect of Magnetic Impurities in a Two-Band Superconductor: A
  Point-Contact Study of Mn-Substituted MgB$_2$ Single Crystals}},}\ }\href
  {\doibase 10.1103/PhysRevLett.97.037001} {\bibfield  {journal} {\bibinfo
  {journal} {Phys. Rev. Lett.}\ }\textbf {\bibinfo {volume} {97}},\ \bibinfo
  {pages} {037001} (\bibinfo {year} {2006})}\BibitemShut {NoStop}%
\bibitem [{\citenamefont {Kumar}\ and\ \citenamefont {Sinha}(1968)}]{kumar}%
  \BibitemOpen
  \bibfield  {author} {\bibinfo {author} {\bibfnamefont {N.}~\bibnamefont
  {Kumar}}\ and\ \bibinfo {author} {\bibfnamefont {K.~P.}\ \bibnamefont
  {Sinha}},\ }\bibfield  {title} {\enquote {\bibinfo {title} {Possibility of
  photoinduced superconductivity},}\ }\href {\doibase 10.1103/PhysRev.174.482}
  {\bibfield  {journal} {\bibinfo  {journal} {Phys. Rev.}\ }\textbf {\bibinfo
  {volume} {174}},\ \bibinfo {pages} {482--488} (\bibinfo {year}
  {1968})}\BibitemShut {NoStop}%
\bibitem [{\citenamefont {Abo-Shaeer}\ \emph {et~al.}(2005)\citenamefont
  {Abo-Shaeer}, \citenamefont {Miller}, \citenamefont {Chin}, \citenamefont
  {Xu}, \citenamefont {Mukaiyama},\ and\ \citenamefont {Ketterle}}]{coh_mol}%
  \BibitemOpen
  \bibfield  {author} {\bibinfo {author} {\bibfnamefont {J.~R.}\ \bibnamefont
  {Abo-Shaeer}}, \bibinfo {author} {\bibfnamefont {D.~E.}\ \bibnamefont
  {Miller}}, \bibinfo {author} {\bibfnamefont {J.~K.}\ \bibnamefont {Chin}},
  \bibinfo {author} {\bibfnamefont {K.}~\bibnamefont {Xu}}, \bibinfo {author}
  {\bibfnamefont {T.}~\bibnamefont {Mukaiyama}}, \ and\ \bibinfo {author}
  {\bibfnamefont {W.}~\bibnamefont {Ketterle}},\ }\bibfield  {title} {\enquote
  {\bibinfo {title} {Coherent molecular optics using ultracold sodium
  dimers},}\ }\href {\doibase 10.1103/PhysRevLett.94.040405} {\bibfield
  {journal} {\bibinfo  {journal} {Phys. Rev. Lett.}\ }\textbf {\bibinfo
  {volume} {94}},\ \bibinfo {pages} {040405} (\bibinfo {year}
  {2005})}\BibitemShut {NoStop}%
\end{thebibliography}%

\end{document}